\documentclass[twocolumn,showpacs,amsmath,amssymb,superscriptaddress]{revtex4}
\usepackage{dcolumn}
\usepackage{color}
\usepackage[dvips]{graphicx}
\usepackage{amsmath}
\usepackage{bm}
\usepackage{url}

\begin{document}

\title{
Shape evolution and the role of intruder configurations in Hg isotopes
within the interacting boson model based on a Gogny energy density functional
}

\author{K.~Nomura}
\email{nomura@ikp.uni-koeln.de}
\affiliation{Institut f\"ur Kernphysik, Universit\"at zu K\"oln, D-50937
K\"oln, Germany}

\author{R.~Rodr\'iguez-Guzm\'an}
\affiliation{Department of Physics and Astronomy, Rice University, Houston, Texas 77005,
USA} 
\affiliation{Department of Chemistry, Rice University, Houston, Texas 77005, USA}

\author{L.~M.~Robledo}
\affiliation{Departamento de F\'\i sica Te\'orica, Universidad
Aut\'onoma de Madrid, E-28049 Madrid, Spain}

\date{\today}

\begin{abstract}
The interacting boson model with configuration mixing, with parameters derived from the
 self-consistent mean-field calculation employing the microscopic Gogny
 energy density functional, is applied to the systematic analysis of the
 low-lying structure in Hg isotopes. 
Excitation energies, electromagnetic transition rates, deformation
 properties, and ground-state properties of the
 $^{172-204}$Hg nuclei are obtained by mapping the microscopic
 deformation energy surface onto the equivalent IBM Hamiltonian in the
 boson condensate. 
These results point to the overall systematic trend of the transition from the 
	   near spherical vibrational state in lower-mass Hg nuclei
	   close to $^{172}$Hg, onset of intruder prolate 
	   configuration as well as the manifest prolate-oblate shape
	   coexistence around the mid-shell nucleus $^{184}$Hg, weakly
	   oblate deformed structure beyond $^{190}$Hg up to the
	   spherical vibrational structure toward the near semi-magic
	   nucleus $^{204}$Hg, as observed experimentally. The quality
 of the present method in the description of the complex shape dynamics
 in Hg isotopes is examined. 
\end{abstract}



\maketitle

\section{Introduction}

Both shape transition and  shape coexistence in
finite nuclei have been a theme of major interests in the field
of low-energy nuclear structure study. 
In particular, significantly low-lying excited
$0^{+}$ states close in energy to the $0^{+}$ ground-state  have
been observed in some nuclei, that reveal the coexistence of different intrinsic shapes. 
Numerous efforts have already been made to better 
understand the nature of such stunning shape phenomena
from both the theoretical and experimental sides 
(see Refs.~\cite{heyde83,wood92,andre00,julin01rev,heyde11} for review). 

In terms of the nuclear shell model 
\cite{federman77,duppen84,heyde85,heyde87,heyde88,heyde95}, the emergence
of low-lying
excited $0^{+}$ states can be  traced back to 
 multiparticle-multihole excitations. 
This scenario applies to the neutron-deficient nuclei in  the Lead region with 
$Z \approx 82$. In this case, two or four protons are excited from the $Z=50-82$ major
shell across the $Z=82$ closed shell to the $h_{9/2}$ orbit.
The residual interaction between the valence protons and neutrons
becomes subsequently enhanced, leading to the lowering of the excited
$0^{+}$ energies. 
This effect is most significant around the neutron mid-shell
$N=104$. 

From the experimental side, extensive $\gamma$-ray spectroscopic studies 
have opened up a vast opportunity to extend the knowledge of the precise
low-lying structure of neutron-deficient Hg isotopes
(cf. \cite{julin01rev,heyde11} for review). 
As observed \cite{julin01rev}, the second $0^{+}$ energy level 
becomes noticeably lower as a function of the neutron
number, starting from around the $^{188}$Hg down to 
the middle of the major shell $N=104$. 
This $0^{+}_{2}$ excited state reaches a minimum in energy around $^{182}$Hg and then  goes up from 
$^{180}$Hg to $^{178}$Hg \cite{elseviers11}. 
The energy levels of the yrast $2^{+}$, $4^{+}$ and $6^{+}$ states
become much higher in $^{172,174,176}$Hg, suggesting the transition to
spherical vibrational states \cite{julin01rev}. 
This picture is quite vividly observed in the parabolic trend of the states, which may belong
to the band built on the low-lying excited $0^{+}$ state, as functions of the neutron number in
the Hg chain. 
On the other hand, from around $^{192}$Hg to 
the
heavier isotopes, the observed energy
levels of the yrast band remain almost constant toward the $N=126$ 
closed shell and form the deformed rotational band \cite{data}. 
Given the recent advances in the experimental studies, it is quite timely, as
well as significant, to address, through a theoretical description of the 
relevant spectroscopy comparable to the experiments, the important
issues of 
the origin of the low-lying excited $0^{+}$ state and the corresponding
shape dynamics in the neutron-deficient Hg isotopes. 

Let us stress, that these heavy mass systems are currently beyond the reach of
large-scale shell-model studies with a 
realistic configuration space and nucleon-nucleon interaction. Therefore a drastic truncation 
scheme is required to make the problem more feasible. 
Such a framework is provided by the interacting boson model (IBM)
\cite{IBM}, which employs the monopole $s$ and  quadrupole $d$ bosons, associated to the $J^{\pi}=0^{+}$
and $2^{+}$ collective pairs of valence nucleons, respectively \cite{Arima77,OAIT,OAI}. 
The collective levels and the transition rates are generated by the 
diagonalization of the boson Hamiltonian composed of only a few
essential interaction terms. 
To handle the configuration mixing in the IBM framework, Duval and 
Barrett proposed to extend the boson Hilbert space to the direct sum of
the configuration sub-spaces corresponding to the $2np$-$2nh$ excitation
($n=0,1,2\ldots$) that comprises $N_{B}+2n$ bosons \cite{duval81}. 
Various features relevant to the 
shape coexistence phenomena in the Pb region have been investigated
within the IBM configuration mixing model: the empirical collective structures from Po down to Pt
isotopes \cite{duval82,Barfield83,decoster99,Fossion03,Garcia11}, geometry and
phases \cite{frank04,frank06,morales08}, and the 
algebraic features (in terms of the so called {\it intruder spin}) \cite{heyde92,DeCoster96,Lehmann97}. 
Nevertheless, one of the main difficulties was that, since 
too many parameters are involved in this
prescription, one has to specify the form of the model Hamiltonian rather {\it a priori} and/or
simply select the parameters by the fit to available data. 

On the other hand, the energy density functional (EDF) framework has been successful in 
the self-consistent mean-field study of bulk nuclear  properties and
collective excitations \cite{Ben03rev} with various classes of effective
interactions, e.g., Skyrme \cite{Skyrme,VB}, Gogny \cite{Gogny}, and those used within relativistic mean-field
(RMF) models \cite{vre05rev,Nik11rev}. 
The mean-field approximation provides the coexisting minima in the 
deformation energy surface, which are associated to the different
intrinsic geometrical shapes \cite{Naza93}. 
Deformation properties and collective excitations, relevant to the
shape coexistence in the neutron-deficient Pb and Hg
 isotopes, have been investigated using  Skyrme \cite{Duguet03,Ben04Pb}, 
Gogny \cite{girod82,Delaroche94,Chasman01,Rayner04Pb,egido04,moreno06} and the RMF interactions \cite{Nik02sc} as well as 
within the Nilsson-Strutinsky method \cite{bengtsson89,Naza93}. 
Recently, a systematic study of the low-lying states in the Lead
region has been performed within the number and angular-momentum
projected generator coordinate method with axial symmetry, employing the
Skyrme EDF \cite{yao13}. 

A method of deriving the
IBM Hamiltonian by combining the density functional framework
with the IBM  has been developed in Ref.\cite{Nom08}. 
Within this method, excitation energies and transition rates 
are calculated by
mapping the deformation energy surface, obtained from the
self-consistent mean-field calculation with a given EDF, onto the
equivalent IBM Hamiltonian in the boson condensate. 
This idea has been applied to the mixing of several multiparticle-multihole
configurations in Lead isotopes \cite{Nom12sc} on the basis of the
Duval-Barrett's technique.

In this paper, we apply the above methodology of deriving the configuration mixing
IBM-2 Hamiltonian parameters from the microscopic Gogny-EDF quantities  to 
the systematic analysis of  low-lying states in a number
of Hg isotopes with mass $A=172-204$, with the focus being on the relevant spectroscopy related to the 
coexistence of different intrinsic shapes around the mid-shell nucleus
$^{184}$Hg.  
The optimal choice of the configuration mixing IBM Hamiltonian consistent with the
EDF-based calculations is identified, and the quality of the procedure
to extract configuration mixing IBM Hamiltonian is addressed. 
We shall use the  parametrization D1M of the Gogny-EDF \cite{D1M}, that 
has  been shown (see, for instance \cite{rayner10odd-1,rayner10odd-2,rayner10odd-3})
to have a similar predictive power in the description of nuclear structure phenomena 
as  the more  conventional D1S \cite{D1S} parameter set.  
Therefore, another motivation of this work is to test the validity of the new
parametrization D1M to the nuclei in Lead region.

This paper is organized as follows: 
our theoretical framework is briefly summarized in Sec.~\ref{sec:framework}.
We then show the microscopic (i.e., EDF) and the mapped
energy surfaces in the considered Hg isotopes in Sec.\ref{sec:pes},
followed by the 
systematic calculations, including the energy levels, deformation properties
(spectroscopic quadrupole moment and the transition quadrupole moment),
and the ground-state properties (mean square charge radii and binding energies) in Sec.~\ref{sec:results}. 
The detailed spectroscopy of selected nuclei exhibiting shape
coexistence is discussed  in Sec.~\ref{sec:results} whereas
Section~\ref{sec:summary} is devoted to the concluding remarks. 
Finally, in appendix \ref{sec:mapping}  
the mapping procedure is described in detail.

\section{Framework\label{sec:framework}}

We first perform a set of constrained 
Hartree-Fock-Bogoliubov (HFB) calculations using the Gogny-D1M
\cite{D1M} EDF to obtain the corresponding mean-field
total energy surface in terms 
of the geometrical quadrupole collective variables $q=(\beta,\gamma)$
\cite{BM}. 
Note that the energy surface in this context denotes the total 
mean-field energy as a function of the deformation variables $q$, where 
neither the mass parameter nor the collective potential is considered explicitly. 
In fact, we only consider the  symmetry-unprojected HFB energy surface
and do not include any zero-point energy corrections.
Having the Gogny HFB energy surface, we subsequently map it onto the
corresponding  IBM energy surface, as described below. 

Turning now to the IBM system, in order to treat the proton cross-shell excitation, we shall use the
proton-neutron IBM (IBM-2) because it is more realistic 
than the original version of the IBM (IBM-1), which does not
distinguish between proton and neutron degrees of freedom. 
The IBM-2 comprises neutron (proton) $s_{\nu}$ ($s_{\pi}$) and
$d_{\nu}$ ($d_{\pi}$) bosons, which reflect the collective
pairs of valence neutrons (protons) \cite{OAI}. 
The number of neutron (proton) bosons, denoted as $N_{\nu}$
($N_{\pi}$), equals half the number of the valence neutrons (protons). 
The doubly-magic nuclei 
$^{164}$Pb and  $^{208}$Pb
are taken as the boson vacua (inert cores). 
As we show below, the Gogny-D1M energy surface exhibits two minima in  $^{176-190}$Hg and
therefore  up to $2p$-$2h$ proton excitations are taken into account to describe 
these nuclei. For the others, i.e.,  $^{172,174}$Hg and $^{192-204}$Hg,
the corresponding energy surfaces exhibit a single mean-field minimum, which is
supposed to be described by a single configuration. 
Therefore, for the nuclei $^{176-190}$Hg, $N_{\pi}$ is fixed, $N_{\pi}=1$ and 3 for the $0p$-$0h$ 
and the $2p$-$2h$ configurations, respectively, while $N_{\nu}$ varies
between 8 and 11. On the other hand, for the nuclei $^{172,174}$Hg and $^{192-204}$Hg, $N_{\pi}=1$
 and $N_{\nu}$ varies between 5 and 6 and between 1 and 7, respectively. 

The IBM Hamiltonian of the system, comprising the normal $0p$-$0h$ and
the $2p$-$2h$ configurations, is written  as
\cite{duval81,duval82} 
\begin{eqnarray}
\label{eq:ham-cm}
 \hat H={\cal\hat P}_{1}\hat H_{1}{\cal\hat
  P}_{1}+{\cal\hat P}_{3}(\hat H_{3}+\Delta_{intr}){\cal\hat
  P}_{3}+\hat H_{mix}, 
\end{eqnarray}
where ${\cal\hat P}_{i}$ ($i=1,3$) stands for the projection operator onto the $N_{\pi}=i$
configuration space. The operator 
$\hat H_{i}$ is the Hamiltonian for the configuration with
$N_{\pi}=i$ bosons 
\begin{eqnarray}
\label{eq:ham}
 \hat H_{i}=\epsilon_{i}\hat n_{d}+\kappa_{i}\hat Q_{\nu}^{\chi_{\nu,i}}\cdot\hat
  Q_{\pi}^{\chi_{\pi,i}}+\kappa^{\prime}_{i}\hat L\cdot\hat L+\sum_{\rho^{\prime}\neq\rho}\hat V_{\rho\rho\rho^{\prime}},
\end{eqnarray}
where the first term $\hat n_{d}=\sum_{\rho}d^{\dagger}_{\rho}\cdot d_{\rho}$ 
($\rho=\nu$ or $\pi$) stands for the
$d$-boson number operator.  The second term in Eq.~(\ref{eq:ham}) 
represents the quadrupole-quadrupole
interaction between the proton and neutron bosons with $\hat
Q_{\rho,i}^{\chi_{\rho,i}}=d^{\dagger}_{\rho}s_{\rho}+s^{\dagger}_{\rho}\tilde
d_{\rho}+\chi_{\rho,i}[d_{\rho}^{\dagger}\times\tilde d_{\rho}]^{(2)}$
being the quadrupole operator. 
The sign of the sum $\chi_{\nu,i}+\chi_{\pi,i}$ specifies whether a
given nucleus is prolate or oblate deformed. 
The third term is 
relevant for rotationally deformed systems, with 
$\hat L=\sqrt{10}\sum_{\rho}[d^{\dagger}_{\rho}\times\tilde
d_{\rho}]^{(1)}$ being the boson angular momentum operator. 
The fourth term on the right-hand side (RHS) of Eq.~(\ref{eq:ham}) stands for the three-body
(cubic) boson term between the proton and neutron bosons  
\begin{eqnarray}
\label{eq:3B}
 \hat V_{\rho\rho\rho^{\prime}}=\sum_{L}
\kappa^{\prime\prime(L)}_{\rho\rho\rho^{\prime},i}
[d^{\dagger}_{\rho}\times
  d^{\dagger}_{\rho}\times d_{\rho^{\prime}}^{\dagger}]^{(L)}\cdot [\tilde
  d_{\rho^{\prime}}\times\tilde d_{\rho}\times\tilde d_{\rho}]^{(L)}, 
\end{eqnarray}
which is identified with the one used in Ref.~\cite{Nom12tri}. 
For each $\rho=\nu$ and $\pi$, there are five linearly independent
combinations in Eq.~(\ref{eq:3B}), identified by the values 
$L=0,2,3,4$ and 6.  In the present study, as in Ref.\cite{Nom12tri}, we only consider
the  $L=3$ term, because its classical limit 
is proportional to $\cos^{2}{3\gamma}$ the only term giving rise to a stable 
triaxial minimum at $\gamma\approx 30^{\circ}$ \cite{Nom12tri}. 
The three-body interaction has been restricted to act only between neutrons and
protons since such proton-neutron correlation becomes more significant in
medium-heavy and heavy nuclei. We have assumed
$\kappa^{\prime\prime(3)}_{\pi\pi\nu,i}=\kappa^{\prime\prime(3)}_{\pi\nu\nu,i}\equiv\kappa^{\prime\prime}_{i}$,
for simplicity \cite{Nom12tri}. 

In Eq.~(\ref{eq:ham-cm}), $\Delta_{intr}$  represents the energy off-set required
to excite two protons from the $Z=50-82$ to the $Z=82-126$ major shells. 
In the same equation, the term $\hat H_{mix}$  stands for the
interaction mixing between the normal and the $2p$-$2h$ configurations
\begin{eqnarray}
\label{eq:ham-mix}
 \hat H_{mix}={\cal\hat P}_{3}(\omega_{s}s^{\dagger}_{\pi}\cdot
  s^{\dagger}_{\pi}+\omega_{d}d^{\dagger}_{\pi}\cdot d^{\dagger}_{\pi}){\cal\hat P}_{1}+h.c., 
\end{eqnarray}
where the parameters $\omega_{s}$ and $\omega_{d}$ are the 
mixing strength between the $N_{\pi}=1$ and the $N_{\pi}=3$ configurations. 

It should be noted that, for each configuration, the Hamiltonian
in Eq.~(\ref{eq:ham}) adopts the simplest possible form, with a minimal number
of parameters consistent with the most relevant topology of the
EDF energy surface.  Up to the third term, the RHS of Eq.~(\ref{eq:ham}) is the standard form
frequently used in a number of IBM-2 calculations \cite{IBM}. In the present study, we include
the so called $\hat L\cdot\hat L$ term since we have a  relatively large
number of bosons, which leads to a deformed rotational spectrum. On the 
other hand, as we will see below, the microscopic energy surface of the considered Hg nuclei
exhibits a triaxial minimum, which requires the inclusion of the cubic 
term in the IBM Hamiltonian \cite{Nom12tri}. 
The physical significance of both the $\hat L\cdot\hat L$ and the cubic terms  has been discussed in
detail in Refs.~\cite{Nom11rot} and \cite{Nom12tri}, respectively. 

The geometrical picture of a given IBM Hamiltonian is provided by the
coherent-state framework \cite{GK}. Such a 
coherent state represents the boson intrinsic wave function
specified by the deformation variables
$\bar q=(\bar\beta_{\nu},\bar\beta_{\pi}, \bar\gamma_{\nu},
\bar\gamma_{\pi})$.  
One can take $\bar\beta_{\nu}=\bar\beta_{\pi}\equiv\bar\beta$ and  
$\bar\gamma_{\nu}=\bar\gamma_{\pi}\equiv\bar\gamma$ as the neutron and
the proton deformations are approximately equal \cite{BM}. 
The  deformation $\bar\beta$ is
assumed to be proportional to the one obtained within the HFB approximation while
$\gamma$ has been
taken to be the same 
in both the HFB and IBM frameworks \cite{GK}. 

In the IBM configuration mixing calculation  one needs to consider the
direct sum of the coherent state for the configuration 
with $N_{\pi,i}=i$ proton bosons \cite{frank04}, denoted here as $|\Phi_{i}(N_{\pi,i},\beta,\gamma)\rangle$. 
The energy surface is obtained as the lower eigenvalue of the $2\times 2$ 
coherent-state matrix \cite{frank04}: 
\begin{eqnarray}
\label{eq:pes-cm}
E(\beta,\gamma)
=\left(
\begin{array}{cc}
E_{11}(\beta,\gamma) & E_{31}(\beta) \\
E_{13}(\beta) & E_{33}(\beta,\gamma)+\Delta_{intr} \\
\end{array}
\right). 
\end{eqnarray}
The diagonal matrix element 
$E_{ii}(\beta,\gamma)=\langle\Phi_{i}(N_{\pi,i},\beta,\gamma)|\hat
H_{i}|\Phi_{i}(N_{\pi,i},\beta,\gamma)\rangle$ on the RHS of Eq.~(\ref{eq:pes-cm})
is given by 
\begin{eqnarray}
\label{eq:pes-diag}
&& E_{ii}(\beta,\gamma)
=
\frac{\epsilon^{\prime}_{i}(N_{\nu}+N_{\pi,i})\bar\beta_{i}^{2}}{1+\bar\beta_{i}^{2}}
+\kappa_{i}N_{\nu}N_{\pi,i}\frac{\bar\beta_{i}^{2}}{(1+\bar\beta_{i}^{2})^{2}}\nonumber \\
&\times&\Big[
4-2\sqrt{\frac{2}{7}}(\chi_{\nu,i}+\chi_{\pi,i})\bar\beta_{i}\cos{3\gamma}+\frac{2}{7}\chi_{\nu,i}\chi_{\pi,i}\bar\beta_{i}^{2}
\Big] \nonumber \\
&-&\frac{1}{7}\kappa^{\prime\prime}_{i}N_{\nu}N_{\pi,i}(N_{\nu}+N_{\pi,i}-2)
\frac{\bar\beta_{i}^{3}}{(1+\bar\beta_{i}^{2})^{3}}\sin^{2}{3\gamma},
\end{eqnarray}
where $\epsilon_{i}^{\prime}=\epsilon_{i}+6\kappa^{\prime}_{i}$ and 
$\bar\beta_{i}\equiv C_{\beta,i}\beta$ with $C_{\beta,i}$ being the 
proportionality coefficient. 
The non-diagonal matrix element is given by $E_{ii^{\prime}}(\beta,\gamma)=\langle\Phi_{i^{\prime}}(N_{\pi,i^{\prime}},\beta,\gamma)|\hat
H_{mix}|\Phi_{i}(N_{\pi,i},\beta,\gamma)\rangle$ ($i\neq
i^{\prime}$), and reads  
\begin{eqnarray}
\label{eq:pes-nondiag}
 E_{13}(\beta)&=&E_{31}(\beta)\nonumber \\
&=&
\sqrt{(N_{\pi,1}+1)N_{\pi,3}}
\Big(
\frac{\omega_{s}+\omega_{d}\bar\beta_{3}^{2}}{1+\bar\beta_{3}^{2}}
\Big)
\nonumber \\
&\times&
\Big(
\frac{1+\bar\beta_{1}\bar\beta_{3}}{\sqrt{(1+\bar\beta_{1}^{2})(1+\bar\beta_{3}^{2})}}
\Big)^{N_{\nu}+N_{\pi,1}}. 
\end{eqnarray}
We also assume hereafter $\omega_{s}=\omega_{d}=\omega$ for simplicity. 

The parameters $\epsilon_{i}^{\prime}$, $\kappa_{i}$, $\chi_{\nu,i}$, 
$\chi_{\pi,i}$, $\kappa_{i}^{\prime\prime}$ for the two independent
IBM-2 Hamiltonians, the energy offset $\Delta_{intr}$, and the mixing strength $\omega$ are
determined following the procedure of Ref.~\cite{Nom12sc} for
the Lead isotopes having three mean-field minima, where the 
approximate separation of the coexisting mean-field minima was assumed. 
Since the procedure to determine all parameters is somewhat lengthy, we
summarize it in Appendix \ref{sec:mapping} in order not to interrupt the
major discussion of the paper. 

As we show below, since the oblate HFB minimum occurs always at smaller deformation
$\beta$ ($\approx 0.15$) than 
the prolate one ($\beta\approx 0.25-0.3$) in most of the nuclei
exhibiting two minima, the Hamiltonians for the $0p$-$0h$ 
and $2p$-$2h$ configurations are associated to the oblate and the
prolate minima, respectively. 
On the other hand, the locations of the oblate and the prolate minima on the $\beta$ axis
remain almost unchanged for the considered nuclei. 
As a consequence, the scale factors $C_{\beta,i}$ remain almost constant, 
i.e., $C_{\beta,1}\approx 3$ and $C_{\beta,3}\approx 5$ for $^{176-190}$Hg. 
For nuclei with a single configuration (i.e., $^{172,174}$Hg and
$^{192-204}$Hg) the $C_{\beta,1}$ becomes larger
as the $N=82$ or 126 closed shells are approached. This is a consequence
of the decreasing  number of valence bosons and the displacement of the 
minimum  towards the origin $\beta=0$. 

On the other hand, a further step is necessary to fix the coefficient of
the $\hat L\cdot\hat L$ term $\kappa^{\prime}_{i}$ as this term only contributes to the
energy surface in the same way as the $\hat n_{d}$ term but 
with a different coefficient $6\kappa^{\prime}_{i}$ (cf. Eq.~(\ref{eq:pes-diag})). 
By following the procedure of
\cite{Nom11rot}, we derive the $\kappa^{\prime}_{i}$ values so that the
cranking moment of inertia for the boson intrinsic 
state \cite{Schaaser86}, which is calculated at the minimum for each unperturbed
configuration with the parameters $\epsilon_{i}^{\prime}$, $\kappa_{i}$,
$\chi_{\nu,i}$, $\chi_{\pi,i}$, $\kappa_{i}^{\prime\prime}$ and
$C_{\beta,i}$ already fixed by the energy-surface mapping, becomes identical to the Thouless-Valatin moment of inertia
\cite{TV} at its corresponding minimum on the HFB energy surface.


With all the  parameters required for an individual nucleus at hand, the
Hamiltonian in Eq.~(\ref{eq:ham-cm}) is diagonalized in the enlarged
model space consisting of the $0p$-$0h$ and the $2p$-$2h$
configurations in the boson $m$ scheme. 
This gives the energy spectra and wave functions for the excited
states, that can be used to compute other properties, as discussed  in Sec.~\ref{sec:results}. 

\section{Energy surfaces\label{sec:pes}}

\begin{figure*}[ctb!]
\begin{center}
\includegraphics[width=17cm]{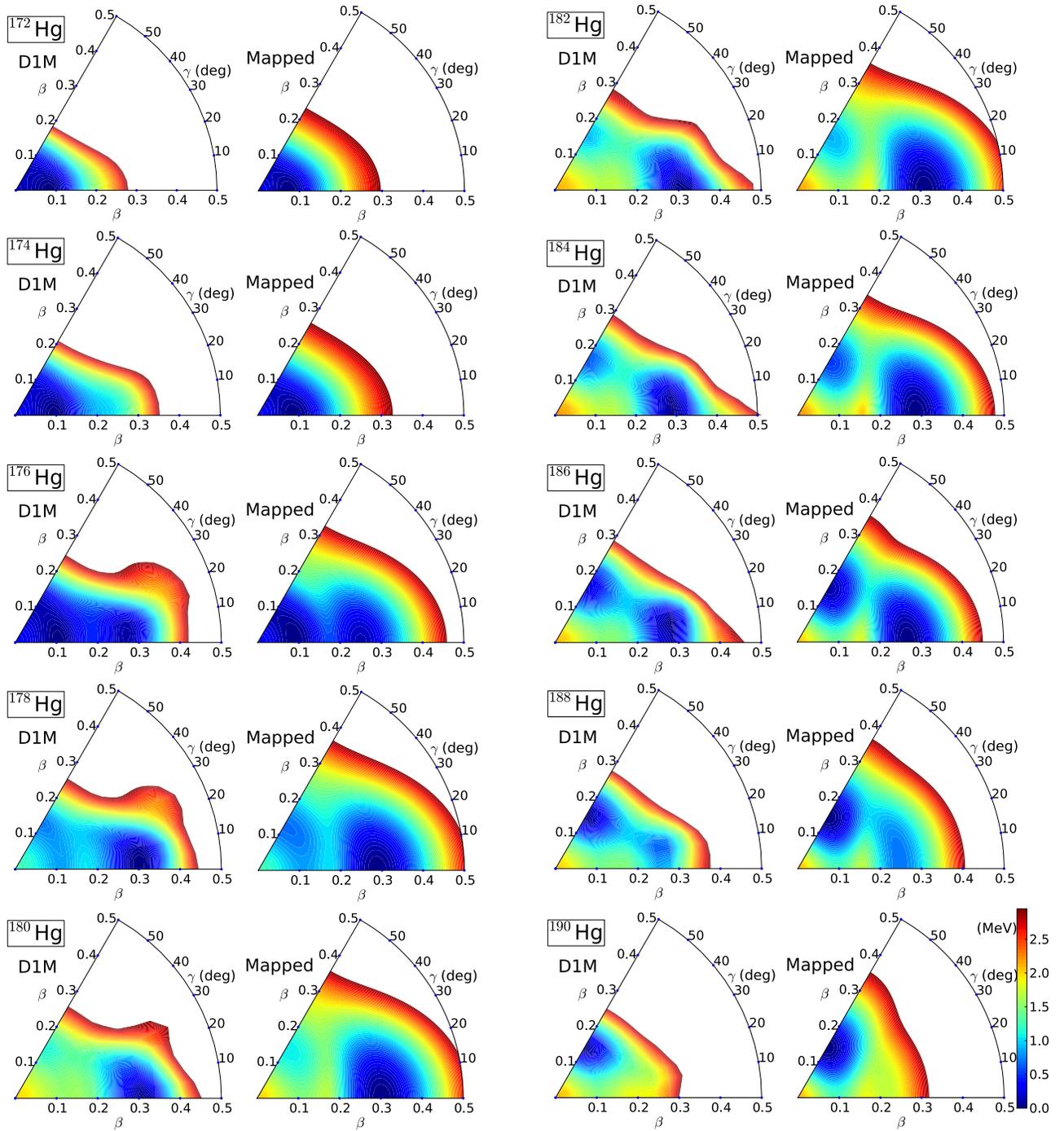}
\begin{tabular}{cccc}
\end{tabular}
\caption{(Color online) Microscopic (``D1M'') and mapped (``Mapped'') potential energy surfaces for the  isotopes $^{172-190}$Hg  in the
 ($\beta,\gamma$)-plane are plotted up to 3 MeV from the absolute
 minimum. The microscopic results are obtained with the Gogny-D1M EDF.
}
\label{fig:pes1}
\end{center}
\end{figure*}

\begin{figure*}[ctb!]
\begin{center}
\includegraphics[width=17cm]{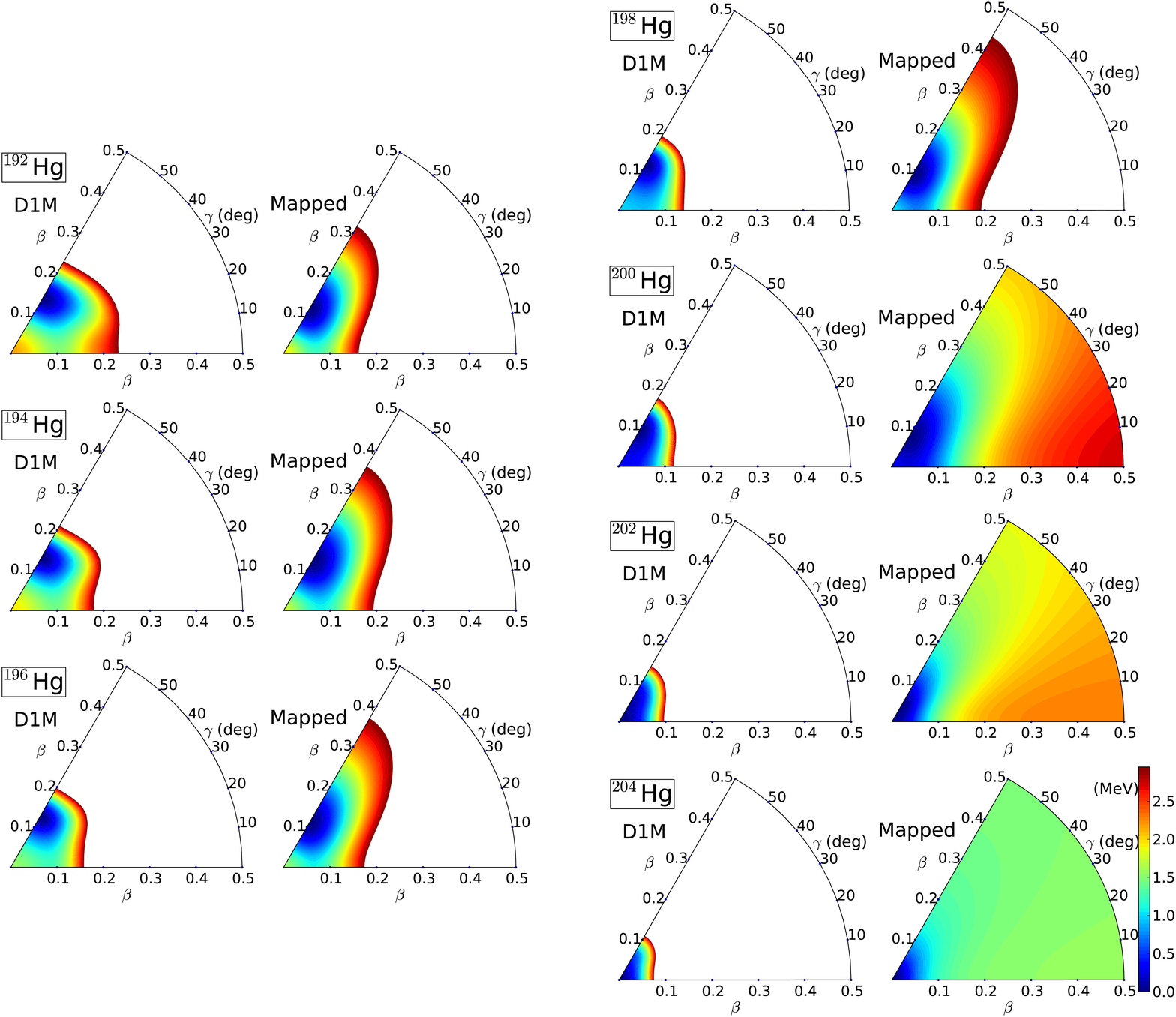}
\begin{tabular}{cccc}
\end{tabular}
\caption{(Color online) Same as Fig.~\ref{fig:pes1}, but for $^{192-204}$Hg. }
\label{fig:pes2}
\end{center}
\end{figure*}

The Gogny-D1M and the mapped energy surfaces of the $^{172-204}$Hg
nuclei are plotted in Figs.~\ref{fig:pes1} and \ref{fig:pes2} in terms of the
$q=(\beta,\gamma)$ deformations. 
We have restricted the plots to configurations up to 3 MeV from the global minimum.
Both the microscopic and the mapped energy surfaces give $\beta_{2}$ values consistent 
with earlier calculations such as  the Nilsson-Strutinsky method
\cite{Naza93} and the collective model approach based on the Gogny-D1S EDF \cite{Delaroche94}, 
where $\beta_{2}\approx -0.15$ and 0.25-0.3 for the oblate and
prolate configurations, respectively. 
Overall, for each individual nucleus, the topology of the mapped IBM energy surface looks 
rather similar to the Gogny-D1M one. They both also follow similar 
systematic changes as a function of the  neutron number, as expected.

Starting from $^{172}$Hg in Fig.~\ref{fig:pes1}, one sees  a nearly spherical
 structure with a
weakly deformed prolate configuration in both $^{172,174}$Hg.
The energy surface suddenly becomes softer along the  $\gamma=0^{\circ}$ axis
from $^{174}$Hg to $^{176}$Hg. 
The latter shows two minima with energies within a
range of  $\approx$ 120 keV. 
Both minima in  $^{176}$Hg  are prolate with
$\beta\approx 0.1$ and  $\beta\approx 0.25$, respectively.  
On the other hand, the prolate minimum with 
$\beta\approx 0.3$ becomes more pronounced in 
$^{178}$Hg while a second one  appears in the oblate
side  with $\beta\approx 0.13$.

The Gogny-D1M energy surfaces for both  $^{180,182}$Hg display more developed 
prolate minima at $\beta\approx 0.3$ while the oblate minimum
at $\beta\approx 0.15$ becomes gradually lower in energy when approaching
$^{184}$Hg for which the energy surface exhibit a  softer
$\gamma$-behavior. Within the HFB 
approximation, the prolate-oblate energy difference in the energy surface reaches a minimum in 
$^{186}$Hg which signals the most prominent case of shape coexistence in the considered  Hg chain.
Note that, in $^{186}$Hg, the higher minimum on the prolate side is a bit off 
the $\gamma=0^{\circ}$ axis, locating at $\gamma\approx 5^{\circ}$. 
In $^{188}$Hg, one can clearly see that the oblate minimum becomes 
energetically favoured over the one around $\gamma\approx 10^{\circ}$ 
on the prolate side. In $^{190}$Hg  only the oblate minimum survives. 

For the  heavier nuclei with $A\ge 192$ in Fig.~\ref{fig:pes2}, the prolate minimum
diminishes and only the oblate one is seen in $^{194-196}$Hg. 
This single oblate minimum becomes softer for $A\ge 198$, and approaches $\beta=0$. 
This implies a structural change from weakly oblate deformed
to nearly spherical states. We have also found an  
almost pure spherical minimum in $^{204}$Hg. It should be noted that the 
corresponding mapped IBM energy surfaces for $^{198-204}$Hg 
look rather flat when compared with the Gogny-D1M ones. This is 
a consequence of 
the limited valance space in these nuclei, close to the shell closure $N=126$, which 
is not large enough to reproduce the topology of the configurations with energies
$\Delta E\ge 1$ MeV. Therefore, we have considered an energy range of up to 1 MeV for
the IBM description of the nuclei $^{198-204}$Hg.

The present calculations, based on the 
Gogny-D1M EDF, predict the oblate minimum to become  the dominant one 
around $^{188,190}$Hg. This is consistent 
with earlier mean-field calculations based on the 
D1 \cite{girod82} and  D1S \cite{Delaroche94} 
parametrizations of the Gogny-EDF. Similar results have also been found 
using the Skyrme-SLy4 EDF \cite{moreno06}. On the other hand, and at variance 
with earlier studies with a deformed Woods-Saxson potential
\cite{bengtsson89,Naza93}, our calculations predict  prolate deformed ground states
for some of the considered neutron deficient Hg isotopes. Let us stress that 
similar results, i.e., prolate ground states, are predicted with the 
Gogny-D1S parameter set (see compilation of the Gogny-D1S HFB results in \cite{CEA}) as well as with other non-relativistic 
Skyrme \cite{moreno06,bender06,yao13} and relativistic NL3 \cite{Nik02sc}
 parametrizations. In fact, the so-called NL-SC (Shape Coexistence)
 parametrization of the relativistic mean-field Lagrangian, which has been specifically adjusted to 
 describe binding energies, radii and deformation in the Lead region, has 
 been introduced in Ref.\cite{Nik02sc} to account for these problem in the 
 more standard NL3 set. All in all, the most standard relativistic and non-relativistic
 parametrizations, used to compute nuclear properties all over the nuclear 
 chart, seem to predict prolate ground states at least for some of the neutron deficient
 Hg isotopes. 
\begin{figure*}[ctb!]
\begin{center}
\includegraphics[width=16.0cm]{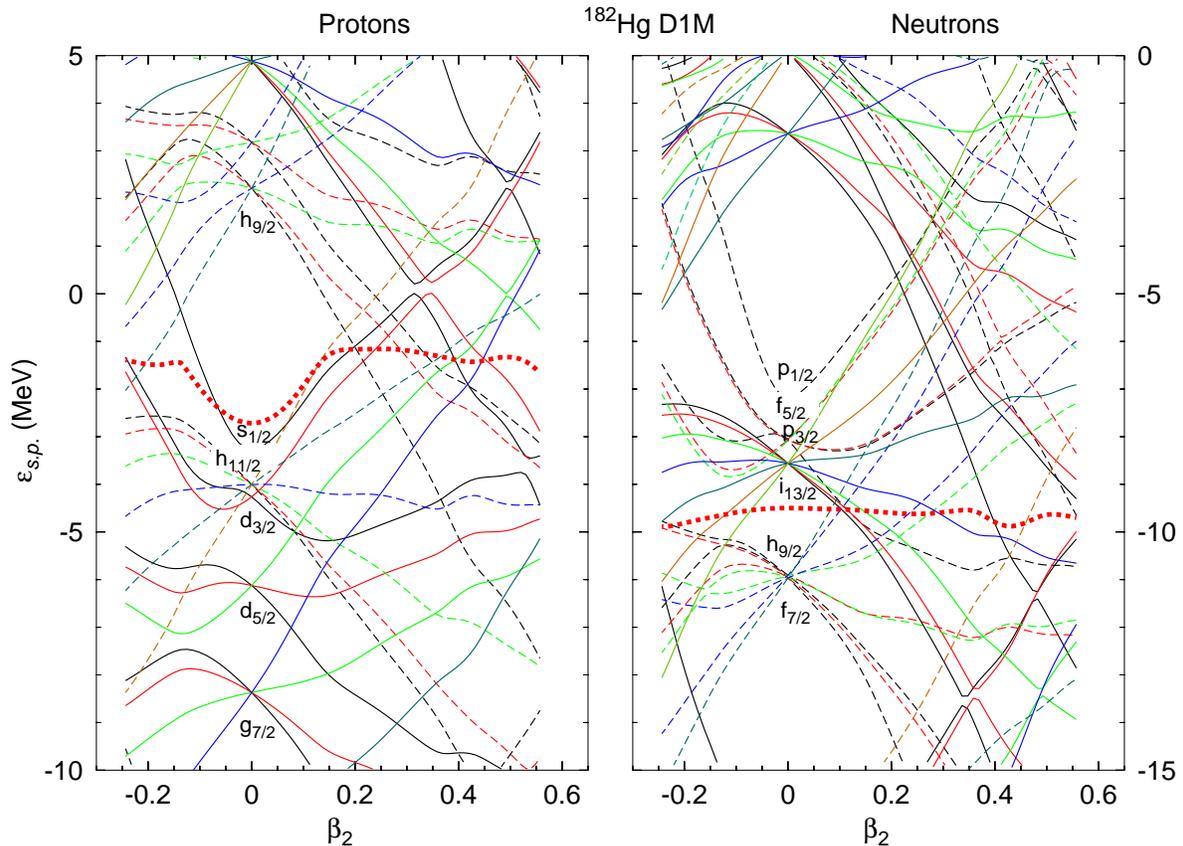}
\caption{(Color online) The single particle energies obtained by diagonalizing
the Hartree- Fock Hamiltonian are plotted as a function of the deformation
parameter $\beta_{2}$ for both protons and neutrons and the nucleus
$^{182}$Hg. Full (dashed) lines correspond to positive (negative) parity orbitals.
The thick dashed line corresponds to the Fermi level. Also the spherical
orbit quantum numbers are given at zero deformation allowing the identification
of each orbital by following its evolution from sphericity.}
\label{fig:182HgSP}
\end{center}
\end{figure*}
In order to clarify the origin of this result we display in Fig \ref{fig:182HgSP} a Nilsson-like plot showing 
 the evolution of the single particle
 energies of the Hartree-Fock Hamiltonian as a function of the axially
 symmetric quadrupole deformation parameter $\beta_{2}$ in the nucleus $^{182}$Hg. The
 choice of this nucleus is guided by its two minima, one oblate and the
 other prolate. 
In the plot, we observe that the deformation of the prolate and
 oblate minima corresponds to the deformation where the proton $h_{9/2}$ 
 and neutron $i_{13/2}$ orbitals cross the Fermi level.  The prolate minimum 
 is to be associated to the crossing of the $K=1/2$ members of the orbitals
 whereas the oblate minimum  to the occupancy of the high-K $K=j$
 members.

\begin{figure*}[ctb!]
\begin{center}
\includegraphics[width=16.0cm]{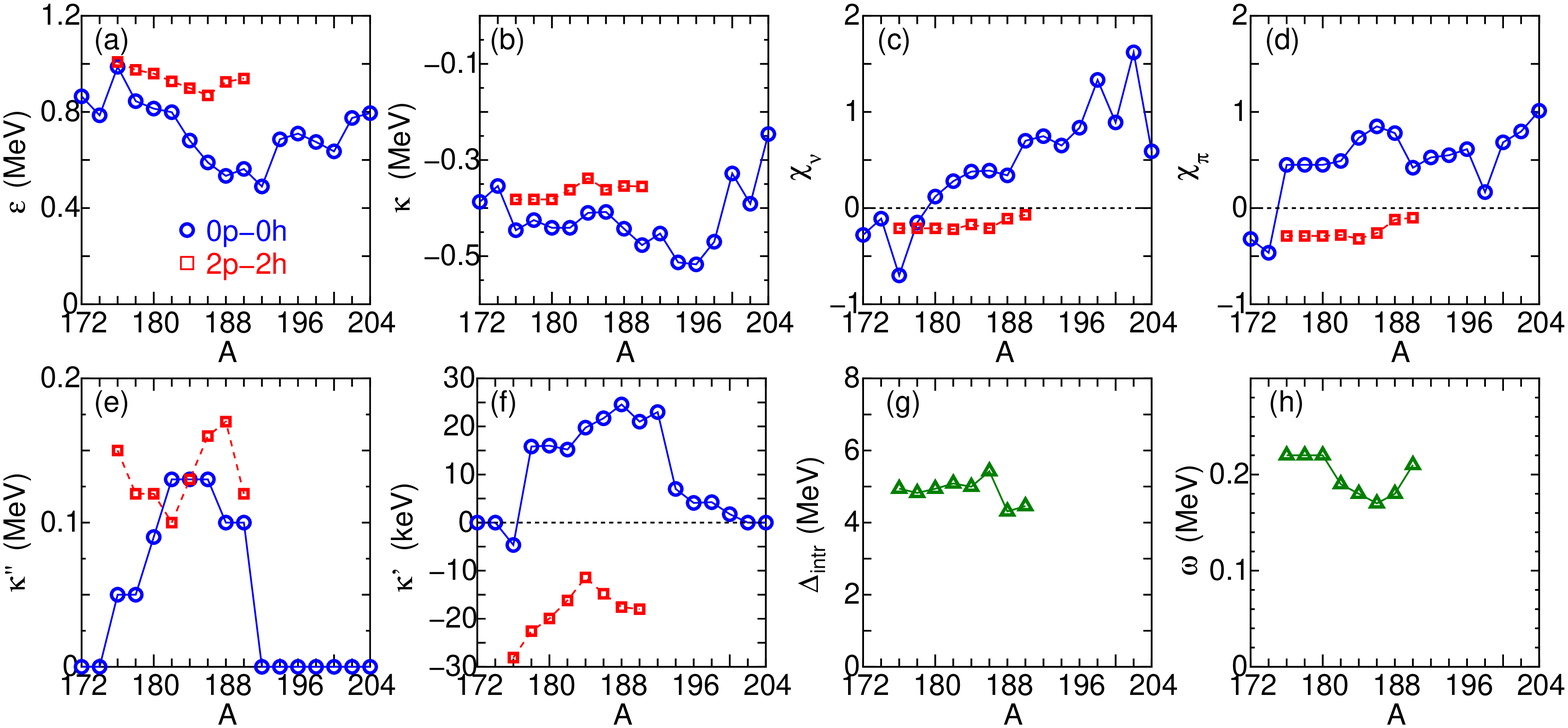}
\caption{(Color online) Derived IBM parameters (a) $\epsilon_{i}$, (b) $\kappa_{i}$,
 (c) $\chi_{\nu,i}$, (d) $\chi_{\pi,i}$, (e) $\kappa_{i}^{\prime\prime}$, (f) $\kappa^{\prime}_{i}$, (g)
 $\Delta_{intr}$ and (h) $\omega$ for the considered $^{172-204}$Hg nuclei as functions
 of mass number $A$. Note that, in panel (f), the parameter $\kappa^{\prime}_{i}$ is
 plotted in keV unit. 
Figure legends in panels (a-f) are indicated in panel (a). }
\label{fig:para}
\end{center}
\end{figure*} 
 
The derived IBM parameters [Eq.~(\ref{eq:ham-cm})] are depicted in Fig.~\ref{fig:para} 
as a function of the  mass number $A$. 
Similarly to its empirical boson number dependence \cite{IBM,OAI}, as well as
to our previous findings \cite{Nom08,Nom10,Nom11sys}, the single $d-$boson energy $\epsilon$ 
shown in panel a) exhibits 
a parabolic behavior centered at mid-shell. This also agrees, with the 
empirical evolution of the $2^{+}$ excited state expected in a given isotope/isotone sequence. 
The $\epsilon_{1}$ and $\epsilon_{3}$ parameters roughly follow this empirical trend. 
Contrary to earlier phenomenological fitting calculations within the IBM
with configuration mixing \cite{duval81,duval82}, the $d$ boson energy
for the intruder configuration 
$\epsilon_{3}$ is always larger than the one for the normal
configuration $\epsilon_{1}$. 
The interaction strengths $\kappa_{1,3}$, shown in panel b), do not change too much. Nevertheless, they
are several times larger than the phenomenological ones 
 ( $\kappa_{1}\approx -0.17\sim -0.14$ MeV 
and $\kappa_{3}\approx -0.14\sim -0.11$ MeV)
\cite{Barfield83}. 
The reason is that the deformation energy, given by the depth of the
minimum in the Gogny-D1M energy 
surface, turns out to be large compared to what is expected from the $\kappa$ value used
phenomenologically. 

We observe in Figs.~\ref{fig:para}(c) and \ref{fig:para}(d) that the sum
$\chi_{\nu}+\chi_{\pi}$ is 
positive (negative) for the oblate (prolate) configuration, 
being consistent with the microscopic energy surface.  
In many of the phenomenological IBM configuration mixing calculations
(e.g., \cite{duval82}), the $2p$-$2h$ configuration has been considered in the 
rotational SU(3) limit of the IBM \cite{IBM} by taking 
$\chi_{\nu,\pi}=-\sqrt{7}/2\approx -1.3$. 
The present result does not follow this trend as both the $\chi_{\nu}$
and $\chi_{\pi}$ values for the $2p$-$2h$ configuration are smaller in
magnitude than the SU(3) limit of $-\sqrt{7}/2$, reflecting a more pronounced 
$\gamma$-soft character for the intruder prolate minimum in the Gogny-D1M energy surface.


From Fig.~\ref{fig:para}(e), one sees that the derived $\kappa^{\prime\prime}_{2}$ value
for both  $^{186}$Hg and $^{188}$Hg is particularly large in agreement with the
Gogny-D1M energy surface [Fig.\ref{fig:pes1}] of the two nuclei displaying 
the most notable $\gamma$ softness on the prolate side in the considered isotopic chain. 
On the other hand, we assume the $\kappa^{\prime\prime}_{1}$ value, for the
single-configuration nuclei $^{172-174,192-204}$Hg, to be zero, because 
neither a triaxial minimum nor notable $\gamma$ softness are observed in the microscopic energy
surfaces shown in  Figs.~\ref{fig:pes1} and \ref{fig:pes2}. 

The $\hat L\cdot\hat L$ coefficient $\kappa^{\prime}_{i}$, shown in panel f),
appears to be stable for the  $0p$-$0h$  configuration 
in $^{178-192}$Hg while a certain decrease in magnitude 
is observed for the $2p$-$2h$ configuration towards the
mid-shell. The sign of $\kappa^{\prime}_{i}$ is of much relevance. 
In particular, a positive (negative) sign for the normal ($2p$-$2h$) configuration
implies that the inclusion of the $\hat L\cdot\hat L$ term reduces
(enlarges) the moment of
inertia of the corresponding unperturbed collective band. 
For the weakly deformed nuclei $^{172-176}$Hg and $^{194-204}$Hg, where only a single configuration
is considered, the derived $\kappa^{\prime}_{1}$ value is almost zero or
very small in magnitude. 

The energy offset $\Delta_{intr}$ in Fig.~\ref{fig:para}(g) changes with
neutron number symmetrically 
with respect to $^{186}$Hg. The
energy needed to excite two protons across the $Z=82$ closed shell becomes maximal for this mid-shell
nucleus because the intruder $2p$-$2h$ configuration gains maximal energy
through deformation. 
As can be observed from panel h), the mixing  strength $\omega$
decreases with boson number toward the midshell.

\section{Results and discussion \label{sec:results}}

In what follows, we compare the results obtained by diagonalizing the mapped 
IBM-2 Hamiltonian with the available experimental data. Results 
for spectroscopic observables, including excitation
energies, $B$(E2) transition rates and  quadrupole moments 
as well as ground-state properties (mean square charge radii and binding energies) are discussed. 


\subsection{Level energy systematics\label{sec:level}}

\begin{figure*}[ctb!]
\begin{center}
\includegraphics[width=17cm]{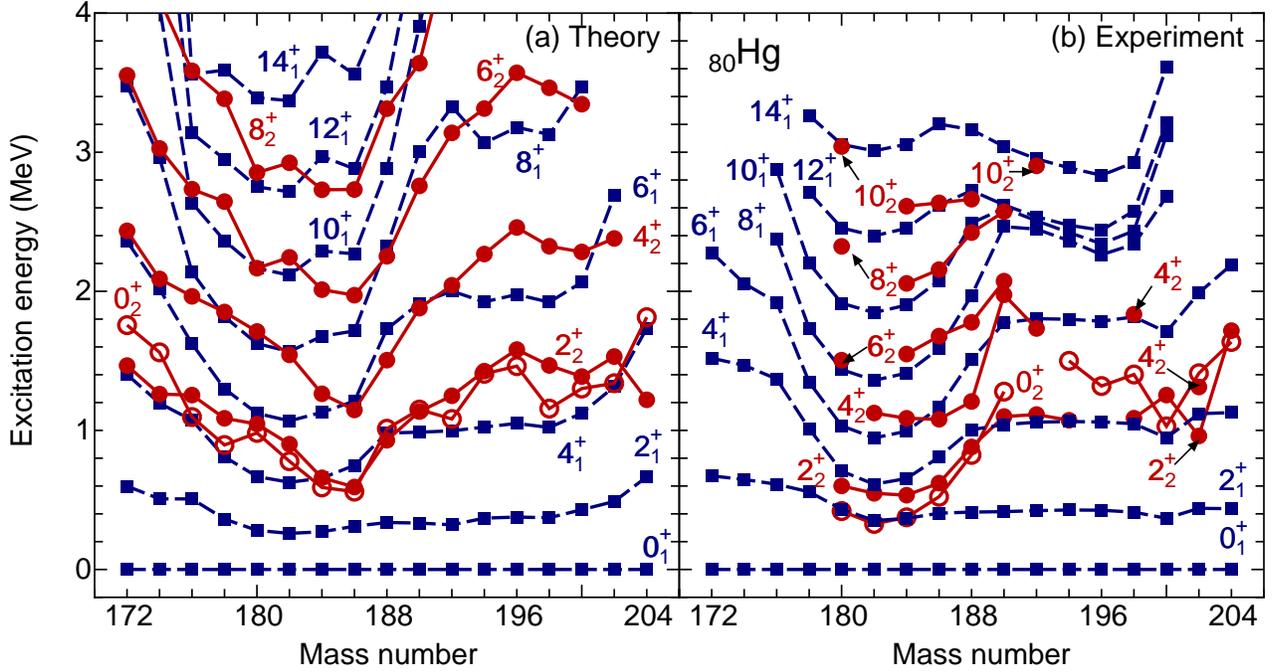}
\caption{(Color online) Level-energy systematics for $^{180-190}$Hg
 isotopes with mass number. Theoretical level energies coming from the mapped IBM-2
 Hamiltonian based on the Gogny-D1M 
 EDF (a) are compared with the experimental \cite{sandzelius09,julin01rev,data,page11,elseviers11} (b) energies. 
 The yrast (square) and the non-yrast (circle) states are
 connected by dashed and solid lines, respectively. 
}
\label{fig:level}
\end{center}
\end{figure*}

The systematics of the excitation energies in the isotopes 
$^{172-204}$Hg is shown in Fig.~\ref{fig:level}. Results 
are presented for states with excitation energies up to 4 MeV. 
The theoretical energy levels, obtained through the diagonalization of the mapped
IBM-2 Hamiltonian, are compared with
the corresponding experimental data \cite{sandzelius09,julin01rev,data,page11,elseviers11}, shown in panel (b). 
As can be seen in Fig.~\ref{fig:level}(a), the calculated spectra for $^{172,174}$Hg
resemble a vibrational-like behaviour  with
$R_{4/2}=E(4^{+}_{1})/E(2^{+}_{1})=$2.34 and 2.36, respectively. We also
observe close lying $4^{+}_{1}$, $2^{+}_{2}$ and $0^{+}_{2}$ levels,
characteristic for the vibrational level structure. 
Although, the excitation energies for the non-yrast states have not been 
experimentally measured, the
experimental $R_{4/2}$ ratios, i.e., 2.26 ($^{172}$Hg) and 2.27 ($^{174}$Hg)
deduced from Fig.~\ref{fig:level}(b), are reproduced well. 
Going from $^{174}$Hg to $^{176,178}$Hg, the $0^{+}_{2}$ level comes down
rapidly in our calculations, being close in energy to the $4^{+}_{1}$
one. This implies that the intruder prolate configuration arises in
$^{176}$Hg as a consequence of the IBM-2 configuration mixing. This 
agrees well with what could be expected from the microscopic energy surface in 
Fig.~\ref{fig:pes1}. 

\begin{figure}[ctb!]
\begin{center}
\includegraphics[width=8.6cm]{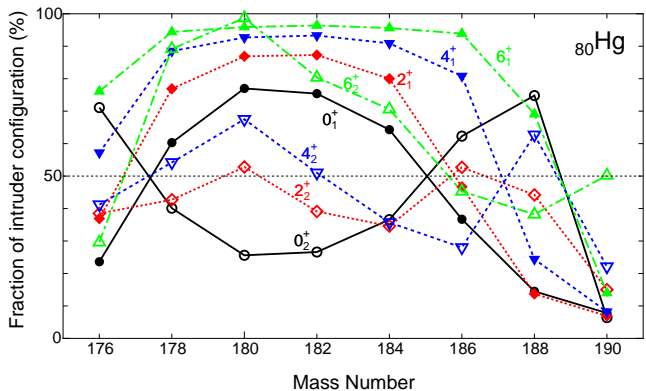}
\caption{(Color online) Fraction of the intruder $2p$-$2h$ configuration in the
 wave functions of the $0^{+}_{1,2}$, the $2^{+}_{1,2}$, the $4^{+}_{1,2}$ and the
 $6^{+}_{1,2}$ excited states of the $^{176-190}$Hg isotopes 
 (in \%). 
}
\label{fig:frac}
\end{center}
\end{figure}

In order to understand the nature of the calculated $0^{+}_{2}$ state for Hg
nuclei with $A\ge 176$, we have calculated 
the probabilities of the different basis states in the wave function of
the state of interest. 
In Fig.~\ref{fig:frac}, we have plotted the fraction of the $2p$-$2h$ component
in the wave functions of the $0^{+}_{1,2}$, $2^{+}_{1,2}$, $4^{+}_{1,2}$
and the $6^{+}_{1,2}$ states. 
The ground-state $0^{+}_{1}$ level for $^{176}$Hg
and $^{178}$Hg are predominantly  $0p$-$0h$ and $2p$-$2h$, respectively. 
On the other hand, the opposite behaviour is predicted for
the $0^{+}_{2}$ state in each of the two nuclei. 
Therefore, the present calculation suggests that the bandhead of the intruder
configuration becomes energetically favoured at $^{178}$Hg over the lowest $0^{+}$
state of the normal configuration. 
Note that, from the energy surfaces in Fig.~\ref{fig:pes1}, both the
$0p$-$0h$ and the $2p$-$2h$ configurations correspond to the prolate
deformation in $^{176,178}$Hg. 

For the $^{180,182}$Hg nucleus in Fig.~\ref{fig:level}(a), however, the level energy of the
$0^{+}_{2}$ state is much higher than the corresponding experimental
data \cite{data,page11,elseviers11}. 
As we will show later, this deviation of the $0^{+}_{2}$ state is mainly
due to the fact that the 
prolate-oblate energy difference is too large in the Gogny-D1M energy
surface. 
Moreover, the present calculation predicts that the ground-state
$0^{+}_{1}$ state in the $^{180,182}$Hg nuclei is comprised mainly of
the intruder prolate configuration. 
Precisely, 77.0\% (75.4\%) of the $0^{+}_{1}$ state of the $^{180}$Hg
($^{182}$Hg) is dominated by the $2p$-$2h$ configuration (see,
Fig.~\ref{fig:frac}). 
However, this contradicts the experimental finding \cite{ulm86,grahn09,elseviers11}
that the ground-state of these nuclei is weakly deformed oblate configuration. 
The reason for the contradiction is that, in the microscopic energy surfaces of
$^{180,182}$Hg (cf. Fig.~\ref{fig:pes1}), the prolate minimum is lower than the oblate one. 

In both $^{184,186}$Hg in Fig.~\ref{fig:level}(a), the excited $0^{+}$ state comes lower in energy, below the
$4^{+}_{1}$ energy level. 
In $^{186}$Hg, in particular, while the $2^{+}_{1}$ level energy is lower 
than the experimental \cite{data} value, the $0^{+}_{2}$ excited state is 
predicted to be the intruder configuration and the oblate bandhead 
becomes the ground state, as shown in Fig.~\ref{fig:frac}. 
Also worth noting is that, similarly to $^{180,182}$Hg, the $^{184}$Hg is predicted to have the prolate
ground state (see Fig.~\ref{fig:frac}) in contradiction with the data \cite{data}. 
From Fig.~\ref{fig:frac}, we notice that the two configurations are strongly
mixed for each of the low-lying states of $^{184,186}$Hg. 
This strong mixing and the subsequent level repulsion may partly account for
the kink observed in the calculated yrast states with $J^{\pi}\ge 10^{+}$ at 
$^{184}$Hg in Fig.~\ref{fig:level}(a).

In accordance with the evolution of deformation in each
configuration, in Fig.~\ref{fig:level}(a) the yrast and the non-yrast states other than $0^{+}_{2}$
states keep lowering toward the mid-shell 
$N=104$, while these levels are generally higher than the experimental
ones in Fig.~\ref{fig:level}(b).

Most of the levels in Fig.~\ref{fig:level}(a) increase their energies when
going from the near mid-shell nuclei $^{184,186}$Hg to $^{188}$Hg, a behaviour
that is consistent with the experimental 
data in Fig.~\ref{fig:level}(b). 
This sudden change in the energy level is also consistent with the Gogny
EDF mean-field energy surface in Fig.~\ref{fig:pes1}, where we observe that 
the intruder prolate minimum becomes less significant in $^{188}$Hg than in
$^{186}$Hg. 
In the ground state of $^{188}$Hg, the oblate normal 
configuration becomes much more populated than the intruder prolate configuration in Fig.~\ref{fig:frac}. 

From Fig.~\ref{fig:frac}, one sees that only a small fraction of the intruder component plays a role 
in the low-lying states of $^{190}$Hg. The excited $0^{+}$ state is 
originated almost purely from a single oblate configuration, consistently with the 
empirical observation \cite{Delaroche94}. 

Back to Fig.~\ref{fig:level}, from $^{190}$Hg to heavier isotopes, the calculated energy
levels of yrast states are almost constant 
as a function of mass number in agreement with the experimental
trend. The energy of most of the non-yrast states keep
increasing as the $N=126$ shell closure is approached. 
In the present model, only the single oblate configuration is required
to describe those nuclei with $192\le A\le 204$. 
A comparison between the experimental and theoretical level structures for
$^{192-200}$Hg in Figs.\ref{fig:level}(a) and \ref{fig:level}(b) reveals 
that, while the signatures of a vibrational-like level distribution ($R_{4/2}$ 
ratio and similar $4^{+}_{1}$,
$2^{+}_{2}$, and $0^{+}_{2}$ energies) suggested experimentally is
roughly reproduced, the calculations suggest a slightly more deformed
rotational character than the experiment. 
This means that the predicted ground-state band is more compressed for the
$2^{+}_{1}$ state but more stretched for $J^{\pi}\ge 6^{+}$ levels than
in the experiment. 
In particular, the theoretical $R_{4/2}$ ratio is generally
$R_{4/2}\approx 2.7\sim 2.8$, while experimental values are 
$R_{4/2}\approx 2.5\sim 2.6$. 

The nuclei $^{200-204}$Hg show deep spherical minima in their energy 
surfaces as seen in Fig.~\ref{fig:pes2}. As a consequence,  
the theoretical $2^{+}_{1}$, $4^{+}_{1}$ and $0^{+}_{2}$
level energies rapidly increase  when the $N=126$ shell closure is approached.
The excitation energies are much larger than the experimental values as 
a consequence of the IBM model space that excludes pure spherical configurations.

To summarize the results in Fig.~\ref{fig:level}(a), we have shown that the present method
describes the shape transition from the near spherical or weakly
deformed structures to the manifest shape coexistence near the mid-shell
nucleus $^{184}$Hg, and to the weakly oblate deformed shapes and the
vibrational structure near the $N=126$ shell closure. 
Although the empirical evidence that the lowest two $0^{+}$ states of
the nuclei around $^{184}$Hg are originated either from the $0p$-$0h$ or the $2p$-$2h$
configurations is reproduced, a major deviation from  experiment
and from earlier phenomenological studies arises in the inverted level structure in 
$^{180,182,184}$Hg and the too compressed $2^{+}_{1}$ level energies overall. 
In the next section, Sec.~\ref{sec:detail}, we address these problems in more
detail.

\subsection{Shape coexistence\label{sec:detail}}

Of all the Hg isotopes,
the nuclei $^{182,184,186,188}$Hg near the neutron mid-shell 
$N \approx104$ are the ones exhibiting the most clear signatures of 
coexistence of different shapes. In order to identify the different
components the level scheme, including both the in-band
and the inter-band E2 transition rates is analyzed. 
To facilitate the comparison with the experimental data, the
excited states shown below will be classified as either 
oblate or prolate bands based on the 
prolate-oblate predominance of the wave function of the state, shown in
Fig.~\ref{fig:frac}, or alternatively on the E2 transition sequence.

The $ B({\rm E2};J\rightarrow J^{\prime})$ transition rate reads
\begin{eqnarray}
  B({\rm E2};J\rightarrow J^{\prime})=\frac{1}{2J+1}|\langle
  J^{\prime}||\hat T^{({\rm E2})}||J\rangle |^{2}, 
\end{eqnarray}
where $|J\rangle$ ($|J^{\prime}\rangle$) represents the wave function of
the initial (final) state with spin $J$ ($J^{\prime}$). 
The E2 operator $\hat T^{\rm (E2)}$ is given as 
$\hat T^{\rm (E2)}=\sum_{\rho,i=1,3}{\cal\hat P}_{i}e_{\rho,i}\hat
Q_{\rho}^{\chi_{\rho,i}}{\cal\hat P}_{i}$, 
where $\hat Q_{\rho}^{\chi_{\rho,i}}$ is identified with the
quadrupole operator in Eq.~(\ref{eq:ham}). We consider the 
same $\chi_{\rho,i}$
value as the one used in the diagonalization of the Hamiltonian in
Eq.~(\ref{eq:ham-cm}). This is equivalent to the so-called
consistent-Q formalism in the IBM-1 \cite{War83}. 
The boson effective charge $e_{\rho,i}$ is assumed to be the same for
protons and neutrons, i.e., $e_{\nu,i}=e_{\pi,i}\equiv e_{i}$. 
In order to obtain an overall systematic agreement with the
typical experimental $B({\textnormal{E2}};2^{+}_{1}\rightarrow 0^{+}_{1})$
value of $\approx 40-60$ Weisskopf units (W.u.), we adopt the values $e_{1}=e_{3}=0.11$ $e$b for $^{182,184}$Hg, $e_{1}=0.07$ $e$b and
$e_{3}=0.15$ $e$b for $^{186}$Hg, $e_{1}=0.15$ $e$b and 
$e_{3}=0.07$ $e$b for $^{176,178,180,188,190}$Hg, and $e_{1}=0.15$ $e$b
for all other nuclei described only with a single configuration.

\subsubsection{$^{182}$Hg}


\begin{figure}[ctb!]
\begin{center}
\includegraphics[width=8.6cm]{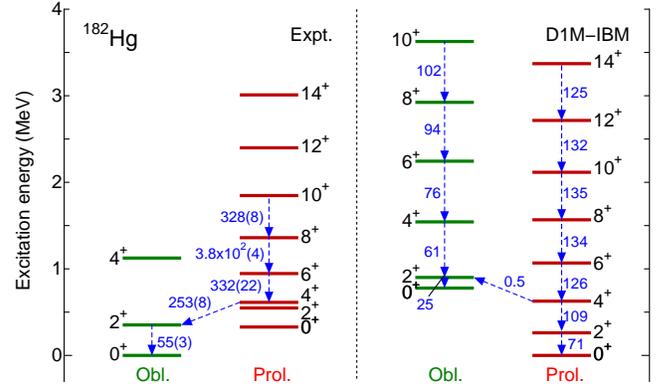}\\
\caption{(Color online) 
Detailed level scheme for the $^{182}$Hg nucleus. The
 level energies and the $B$(E2) values in Weisskopf units (numbers written along arrows)
 obtained from the mapped IBM-2 Hamiltonian based on the Gogny-D1M EDF
 are shown. 
The oblate and the prolate bands are indicated by ``Obl.''
 and ``Prol.'', respectively. 
The experimental data are taken from Ref.~\cite{data}. }
\label{fig:182}
\end{center}
\end{figure}

The detailed level scheme of the nucleus 
$^{182}$Hg is shown in Fig.~\ref{fig:182}. 
The spectra and the $B$(E2) transition
strengths, computed  from the mapped IBM-2 Hamiltonian based on 
the Gogny-D1M EDF, are compared with the relevant experimental
\cite{data} level scheme in the same
figure. 

The calculation predicts the ground-state band  
to have an intruder prolate nature, whereas experimentally the ground
state of $^{182}$Hg has been suggested to be of oblate nature \cite{ulm86}. 
A clear collective pattern is seen from the calculated E2 transition sequence
for both the predicted prolate and oblate bands, while the 
experimental $10^{+}\rightarrow 8^{+}$, $8^{+}\rightarrow 6^{+}$ and $6^{+}\rightarrow 4^{+}$ E2 transition rates in the prolate band are
underestimated in the present calculation. 
Major deviation from the experimental data occurs in the inter-band 
transition from the $4^{+}$ state in the prolate band to the 
$2^{+}$ state in the oblate band. 
The corresponding experimental value, $B({\rm E2};4^{+}_{1}\rightarrow 2^{+}_{1})=253(8)$ W.u,  
is quite large in comparison to 
the $2^{+}_{1}\rightarrow 0^{+}_{1}$ E2 transition strength reflecting the  very
strong mixing between the different configurations. 
In our calculations, however, relative to the 
$B({\rm E2};2^{+}_{1}\rightarrow 0^{+}_{1})=71$ W.u., the inter-band
transition 
$B({\rm E2};4^{+}_{1}\rightarrow 2^{+}_{2})=0.5$ W.u. appears to be too
small, implying that the mixing effect is not significant. 
This could be relevant for the predicted level structure where, in comparison to the
experimental data, the energy levels of the second band 
(which is of oblate nature in the present work) are systematically
higher than those of the ground-state band, giving rise to a weak E2 transition. 
This problem is traced back to the unexpectedly large energy difference
between the prolate and the oblate HFB minima (see, Fig.~\ref{fig:pes1}). 

A similar argument holds for $^{180}$Hg, where the contents of the
two configurations in the wave functions of the low-lying states look
similar to the ones of  $^{182}$Hg (see 
Fig.~\ref{fig:frac}). 

\subsubsection{$^{184}$Hg}

\begin{figure}[ctb!]
\begin{center}
\includegraphics[width=8.6cm]{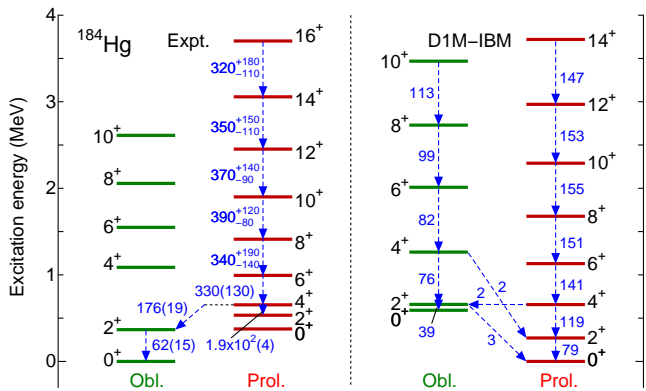}
\caption{(Color online) 
Same as Fig.~\ref{fig:182}, but for $^{184}$Hg. 
The experimental data are taken from Ref.~\cite{data}. 
}
\label{fig:184}
\end{center}
\end{figure}

The results for the $^{184}$Hg nucleus shown in Fig.~\ref{fig:184}  look 
rather similar to the ones obtained  for $^{182}$Hg depicted in Fig.~\ref{fig:182}.
Both show a  prolate ground state and 
the $B$(E2) systematics in each band looks similar. 
In particular, the weak inter-band $B$(E2) transitions between the $4^{+}_{1}$ and the $2^{+}_{2}$ excited
states as well as between $4^{+}_{2}$ and the $2^{+}_{1}$ states
indicate that the mixing between the two configurations may not have a significant effect. 
The reason for this discrepancy seems to be the same as in $^{182}$Hg. 

\subsubsection{$^{186}$Hg}

\begin{figure*}[ctb!]
\begin{center}
\includegraphics[width=15.0cm]{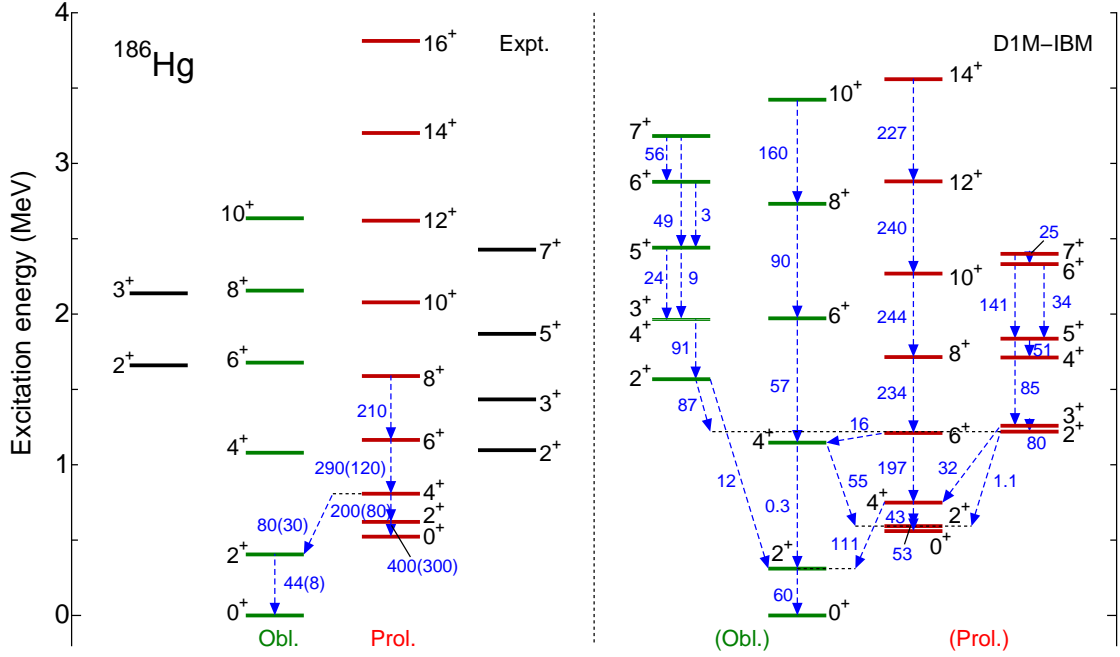}
\caption{(Color online) 
Same as Fig.~\ref{fig:182}, but for $^{186}$Hg
 nucleus. 
The experimental data are taken from Ref.~\cite{data}. 
To help identify the energies of $3^{+}_{2}$ and
 $4^{+}_{4}$ states, closely lying to with other, 
$E(3^{+}_{2})=1.970$, $E(4^{+}_{4})=1.959$ (in MeV), 
and the relevant $B$(E2) transitions are $B({\rm
 E2};3^{+}_{2}\rightarrow 2^{+}_{4})=91$, $B({\rm
 E2};5^{+}_{2}\rightarrow 3^{+}_{2})=9$, $B({\rm
 E2};5^{+}_{2}\rightarrow 4^{+}_{4})=24$ (in W.u.) in the right panel. 
}
\label{fig:186}
\end{center}
\end{figure*}

The empirical prolate-oblate assignment of the lowest two collective
bands in $^{186}$Hg suggests \cite{ma93} that the band built on the $0^{+}_{1}$ state
is of oblate nature while the one on the $0^{+}_{2}$ state is of
prolate. 
Following this assignment, the $0^{+}$ ground-state  
in $^{186}$Hg  (see, Fig.~\ref{fig:186}) 
is predicted to be of oblate nature in the present work. 
The $0p$-$0h$ and the intruder $2p$-$2h$ configurations are substantially mixed in the
wave functions of the lowest two $0^{+}$ states.
According to the results of Fig.~\ref{fig:frac}, the $0^{+}_{1}$ ($0^{+}_{2}$)
state contains 36.7 (62.3)\% of the intruder $2p$-$2h$
configuration. 
In fact, the prolate configuration is quite notable in the mean-field
energy surface shown in Fig.~\ref{fig:pes1}.  
Our calculations seem to follow the 
experimental energy levels with up to $J^{\pi}=4^{+}$ in the oblate
band, the energy levels in the
intruder prolate band, and some of the available $B$(E2) data. 
The deviation is seen in the stretching of the $J^{\pi}=6^{+}_{2}$ and
the $8^{+}_{2}$ energy levels in the oblate band, and in the
$B({\textnormal{E2}};4^{+}_{1}\rightarrow 2^{+}_{2})$ and $B({\textnormal{E2}};2^{+}_{2}\rightarrow 0^{+}_{2})$ in
the prolate band. 
In addition, rather irregular in-band $B$(E2) transitions are found in 
the $4^{+}\rightarrow 2^{+}$ transitions, which are indeed quite weak
compared to the inter-band $4^{+}\rightarrow 2^{+}$ transitions. 
This implies a pronounced mixing between the two configurations. 


We also analyze the structure of higher-lying bands, including odd-spin
states as
there are sufficient experimental data to compare with.  
From Fig.~\ref{fig:186} one sees that the present calculation also
reproduces the excitation energies of odd-spin states rather well. 
The odd-spin states, $3^{+}_{1}$, $5^{+}_{1}$ and $7^{+}_{1}$
($3^{+}_{2}$, $5^{+}_{2}$ and $7^{+}_{2}$),  are 
predicted to be the members of the prolate (oblate) band. 
Due to the strong E2 transition sequence, it is also likely that a set of the states
 $2^{+}_{3}$, $3^{+}_{1}$, $4^{+}_{3}$, $5^{+}_{1}$, $6^{+}_{3}$
and $7^{+}_{1}$ ($2^{+}_{4}$, $3^{+}_{2}$, $4^{+}_{4}$, $5^{+}_{2}$, $6^{+}_{4}$
and $7^{+}_{2}$), forms a quasi-$\gamma$, i.e., $K^{\pi}=2^{+}$, band
for the prolate (oblate) configuration. 
These two quasi-$\gamma$ bands seem to be close in energy, with $2^{+}$ bandheads
being within 400 keV. 
One can also observe in both quasi-$\gamma$ bands quite strong E2 transitions
from the 
$3^{+}$ level to the corresponding $2^{+}$ bandhead, which is
in the same order of magnitude as the $2^{+}\rightarrow 0^{+}$ E2 transition in
each $K^{\pi}=0^{+}$ band, and 
those between the members of the quasi-$\gamma$ band. 
Note that this prediction (i.e., the existence of the two quasi-$\gamma$ bands) is consistent with the empirical assignment of these
levels \cite{Delaroche94}, including the collective model description
based on Gogny-D1S EDF. 
In addition, one should notice that the quasi-$\gamma$ band in the prolate configuration looks
similar to the one predicted within the 
rigid-triaxial rotor model of Davydov
and Filippov, characterized by the the doublets ($2^{+}_{\gamma}$,$3^{+}_{\gamma}$),
($4^{+}_{\gamma}$,$5^{+}_{\gamma}$), 
($6^{+}_{\gamma}$,$7^{+}_{\gamma}$), \ldots etc \cite{Davydov58}. 
Empirically the Davydov-Filippov picture is rarely realized. 
Therefore, the level structure in the proposed quasi-$\gamma$ band of
prolate nature (in Fig.~\ref{fig:186}) seems to be just a consequence of the mixing, which
pushes up the energy levels of the even-spin states in the band.

From the comparison with the IBM phenomenology for $^{186}$Hg
\cite{Barfield83}, we notice that the present result exhibits a similar
level of agreement with the experiment
regarding the energy spectra of the oblate
ground-state band with $J\leq 6^{+}$. 
Our result reproduces  slightly better the 
$3^{+}_{1}$ state while, as discussed in Sec.~\ref{sec:level}, the prolate $0^{+}_{2}$
band-head energy is a bit more overestimated. 

That the oblate band is the ground state and that
the intruder prolate band is built on the $0^{+}_{2}$ state, are  
consistent with the result of the most recent projected GCM calculation
of the $^{186}$Hg nucleus with the Skyrme SLy6 functional \cite{SLy} (see Fig.~16 in
\cite{yao13}). However, there is a certain 
quantitative difference between the two descriptions.

\subsubsection{$^{188}$Hg}


\begin{figure}[ctb!]
\begin{center}
\includegraphics[width=8.6cm]{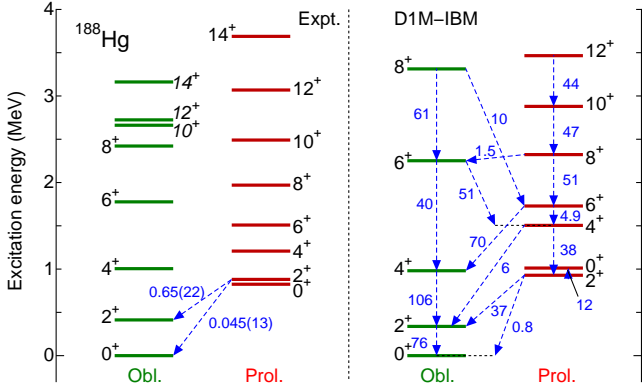}
\caption{(Color online) 
Same as Fig.~\ref{fig:182}, but for $^{188}$Hg nucleus. The experimental data are taken from Ref.~\cite{data}. 
} 
\label{fig:188}
\end{center}
\end{figure}

For the $^{188}$Hg nucleus, the results shown in Fig.~\ref{fig:188} 
show a rather reasonable agreement between theory and experiment 
regarding the band structure and including the energy level of
the $0^{+}_{2}$ state. 
Note that the experimental $10^{+}_{1}$, $12^{+}_{1}$ and $14^{+}$ 
states, written in italic in Fig.~\ref{fig:188}, are assigned to be 
members of a oblate band different from the ground-state oblate band \cite{data}. 
A pronounced  mixing between the two configurations is confirmed from 
the $B$(E2) values of the  $6^{+}_{1,2}\rightarrow 4^{+}_{1,2}$ transitions. 
They reflect the sizable amount
of mixing seen in the wave functions of the two $6^{+}$ excited
states in Fig.~\ref{fig:frac} [the wave function of the
$6^{+}_{1}$ ($6^{+}_{2}$) state contains  69.1 (38.2)\% of the
$2p$-$2h$ component]. 
While the dominance of the $2^{+}_{2}\rightarrow 2^{+}_{1}$ E2 transition
over the $2^{+}_{2}\rightarrow 0^{+}_{1}$ transition roughly follows 
the experimental trend, their absolute values are much larger than the
data. 
The reason is again the very strong mixing between the low-spin states. 
In $^{188}$Hg, any sequence of the 
quasi-$\gamma$ band structure has not been obtained 
in our calculations. 
On the other hand, the predicted $3^{+}_{1}$ energy (1.437 MeV) 
agrees well with the data (1.455 MeV) \cite{data}.

\subsubsection{Mixing matrix element}

\begin{figure}[ctb!]
\begin{center}
\includegraphics[width=7.0cm]{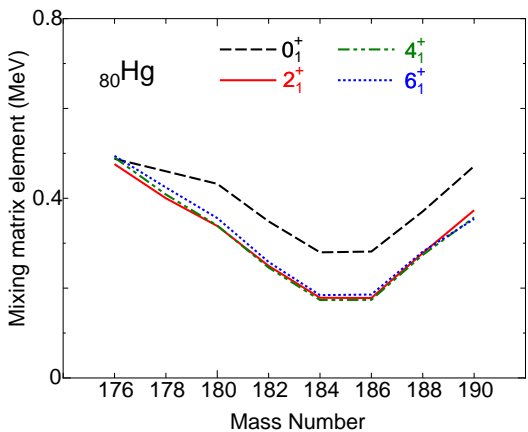}
\caption{(Color online) Matrix element of the mixing interaction $|\langle\hat
 H_{mix}\rangle|$ between the $0^{+}_{1}$, $2^{+}_{1}$, $4^{+}_{1}$ and the $6^{+}_{1}$ excited
 states diagonalized in the unperturbed $0p$-$0h$ configuration space and the
 corresponding states resulting from the diagonalization in the
 unperturbed $2p$-$2h$ configuration space for the nuclei for which
 configuration mixing calculation is performed. }
\label{fig:mix}
\end{center}
\end{figure}
  
In the last two sections we have found that, in the $^{186,188}$Hg nuclei, the mixing between
the two configurations can be too large for the low-spin states, resulting in some discrepancies
with the experimental data. 
To shed some light into the origin of the mixing we display in Fig.~\ref{fig:mix} the matrix
element of the mixing interaction $\langle\hat H_{mix}\rangle$, that couples the
$0^{+}_{1}$, $2^{+}_{1}$, $4^{+}_{1}$ and $6^{+}_{1}$ excited states resulting from
the unperturbed $0p$-$0h$ and $2p$-$2h$ Hamiltonians for $^{176-190}$Hg.  

These results for the mixing matrix elements may explain the calculated
level energy spacings, e.g., between $2^{+}_{1}$ and $2^{+}_{2}$ states in
$^{186}$Hg 
which is larger than the corresponding experimental data,
and also confirm the too strong mixing in these low-spin states. 
To compare with the schematic two-level mixing calculations, the present
$|\langle\hat H_{mix}\rangle_{0^{+}_{1}}|$ value of 281 keV for 
$^{186}$Hg is of the same order of magnitude as the earlier result of
$>110$ keV \cite{joshi94}, but is much larger than a 
more recent result of $69^{+25}_{-41}$ keV \cite{scheck10}. 

For the $^{188}$Hg nucleus, the mixing matrix element $|\langle\hat H_{mix}\rangle|\approx 280$ keV for the
unperturbed $2^{+}_{1}$ and $6^{+}_{1}$ states (cf. Fig.~\ref{fig:mix}) could explain the quenching
of the $2^{+}$ level and the stretching of $J^{\pi}\ge 6^{+}$ level in 
the ground-state band, as shown in Fig.~\ref{fig:188}. 
On the other hand, the mixing matrix element for the unperturbed $0^{+}_{1}$
state for $^{188}$Hg is 371 keV, which seems to be large enough to account for the
calculated $0^{+}_{2}$ excitation energy of 1012 keV. 

\subsubsection{Discussion}

To summarize the results of Sec.~\ref{sec:detail}, we list the main deficiencies
of our model in its current version:
(i) In $^{182}$Hg (as well as $^{184}$Hg), the predicted band structure is in 
contradiction with the empirical assignment, i.e., the ground-state band
is of weakly deformed oblate nature. 
(ii) In $^{184,186,188}$Hg, 
irregular patterns appeared in some of the energy levels
in the oblate band and 
in the $B$(E2) transitions between low-spin states. 
Overall, the $2^{+}_{1}$ energy level has been predicted to have an energy
too low in comparison with the experimental data. 

A major reason for the inversion of the prolate and the oblate
bands in $^{180,182,184}$Hg could be the peculiar topology of
the microscopic energy surface, i.e., the energy difference between
the prolate and oblate mean-field minima. 
This seems to be quite likely since, as we have shown in the
Gogny-D1M energy surfaces e.g., for $^{180}$Hg ($^{182}$Hg) in Fig.~\ref{fig:pes1}, the oblate minimum at 
$\beta\approx 0.15$ looks higher in energy approximately by 1.1 (0.9) MeV than
intruder prolate minimum at $\beta\approx 0.3$. 
This prolate-oblate energy difference is too large to reproduce the energy
spacing of the experimental $0^{+}_{1}$ and $0^{+}_{2}$ levels of
$\approx 400$ keV in $^{180,182}$Hg, and to explain the empirical 
systematics, i.e., the weakly oblate deformed ground-state band. 
On the other hand, as pointed out in Sec.~\ref{sec:pes}, most 
of the standard EDF parameterizations, used for the global description
of the nuclear properties over the whole periodic table, commonly predict the prolate ground state in
the mean-field energy surface. 
We note that, in the recent beyond mean-field calculation on the low-lying structure
in the neutron deficient Hg isotopes \cite{yao13}, the prolate ground
state has been predicted in the $^{180,182,184}$Hg nuclei, similarly to
our results.  

Another reason can be that the  IBM-2 parameters deduced with our method
are not good enough to describe all the
details of the experimental low-lying states. 
For instance, the too strong mixing in the low-spin
states in $^{186,188}$Hg can be traced back to the value for the mixing strength $\omega$
used in this work that perhaps is so large as to make the energy spacing between 
$0^{+}_{1}$ and $0^{+}_{2}$ states larger than experimental data. 
The discrepancies in the $B$(E2) systematics, as well as the stretching of the 
lower band, may arise from the fact that the derived
quadrupole-quadrupole interaction $\kappa_{i}$ is rather large, being
more than twice as large as the one used in the earlier IBM-2 phenomenology 
\cite{duval81,duval82,Barfield83}.  
Such large value of $|\kappa_{i}|$ reflects that the deformation energy,
measured by the depth of the minimum, in the Gogny-D1M is unexpectedly
large. 
This seems to be a common feature for any EDF parametrization, and is
consistent with the conclusions in our previous studies in other
isotopic chains (e.g., \cite{Nom10,Nom11sys}). 
In the case where only a single minimum is concerned, the effect of the
too strong quadrupole-quadrupole interaction could be effectively included in the boson
effective charge, leading to a reasonable agreement with the
experiment \cite{Nom12tri}. 
However, this is not the case with the present work because we consider
much more complex systems with more then one configurations and also
there are so many Hamiltonian parameters and (four) adjustable effective
charges. 
Therefore, it is quite unlikely that a significantly better agreement
with the experimental $B$(E2) systematics could be obtained only by
adjusting the effective charges. 

\subsection{Spectroscopic quadrupole moment}

\begin{figure}[ctb!]
\begin{center}
\includegraphics[width=8.6cm]{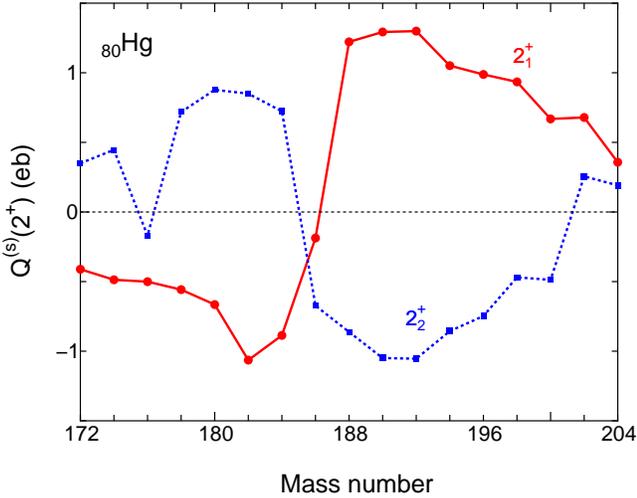}
\caption{(Color online) Calculated spectroscopic quadrupole moments $Q^{(s)}$ for the lowest
 two excited $2^{+}$ states of the considered Hg nuclei as functions
 of mass number. }
\label{fig:qm}
\end{center}
\end{figure}

To confirm from an alternative perspective whether each individual Hg nucleus is oblate or prolate
deformed  
we have also analyzed the spectroscopic quadrupole moments for the lowest two
$2^{+}$ excited states, which belong either to the first oblate or prolate
band for the near mid-shell nuclei. 
The spectroscopic quadrupole moment $Q^{(s)}$ for a state with spin $J$
reads
\begin{eqnarray}
\label{eq:qm}
 Q^{(s)}(J)=\sqrt{\frac{16\pi}{5}}
\left(
\begin{array}{ccc}
J & 2 & J \\
-J & 0 & J \\
\end{array}
\right)
\langle J||\hat T^{({\rm E2})}||J\rangle. 
\end{eqnarray}

The overall systematic trend in $Q^{(s)}(2^{+}_{1,2})$ seems to
correlate well with the evolution of mean-field minima shown in
Figs.~\ref{fig:pes1} and \ref{fig:pes2} and with the structure of the corresponding wave functions in Fig.~\ref{fig:frac}. 

The calculated quadrupole moments for the
$2^{+}_{1}$ and the $2^{+}_{2}$  states 
in all the considered nuclei $^{172-204}$Hg, are shown in 
Fig.~\ref{fig:qm}. 
For the lightest isotopes $^{172,174}$Hg, considered in a single
configuration, $Q^{(s)}(2^{+}_{1})<0$ confirms the prolate deformation
in the ground state. 
At the $^{176}$Hg nucleus, where the two prolate configurations are
considered, both $Q^{(s)}(2^{+}_{1})$ and  $Q^{(s)}(2^{+}_{2})$ are
negative, as expected.  
From $^{178}$Hg to $^{184}$Hg, $Q^{(s)}(2^{+}_{1})<0$
($Q^{(s)}(2^{+}_{2})>0$) gradually increases in magnitude, consistently
with the growing prolate (oblate) minimum in the ground state. 
For $^{186}$Hg, $Q^{(s)}(2^{+}_{2})$ changes its sign as the $2^{+}_{2}$
belongs to the prolate band, while the negative value of $Q^{(s)}(2^{+}_{1})$ 
contradicts the corresponding energy surface in Fig.~\ref{fig:pes1} and
the level scheme in Fig.~\ref{fig:186}, where the ground state is oblate deformed. 
Nevertheless, the magnitude of 
$Q^{(s)}(2^{+}_{1})<0$ for the $^{186}$Hg nucleus is quite small, due to the significant effects of the configuration mixing in
the $2^{+}_{1}$ state (Fig.~\ref{fig:frac}) and  
the triaxiality (Fig.~\ref{fig:pes1}), as compared to
other isotopes. 
From $^{188}$Hg to the heavier isotopes, the predicted $Q^{(s)}$ systematics is in
agreement with the energy surfaces in Figs.~\ref{fig:pes1} and
\ref{fig:pes2}, with $Q^{(s)}(2^{+}_{1})>0$ and $Q^{(s)}(2^{+}_{2})<0$),
indicating the oblate ground-state band, both decreasing in magnitude
as approaching the neutron shell closure $N=126$.

\subsection{Transition quadrupole moment}

\begin{figure*}[ctb!]
\begin{center}
\includegraphics[width=14cm]{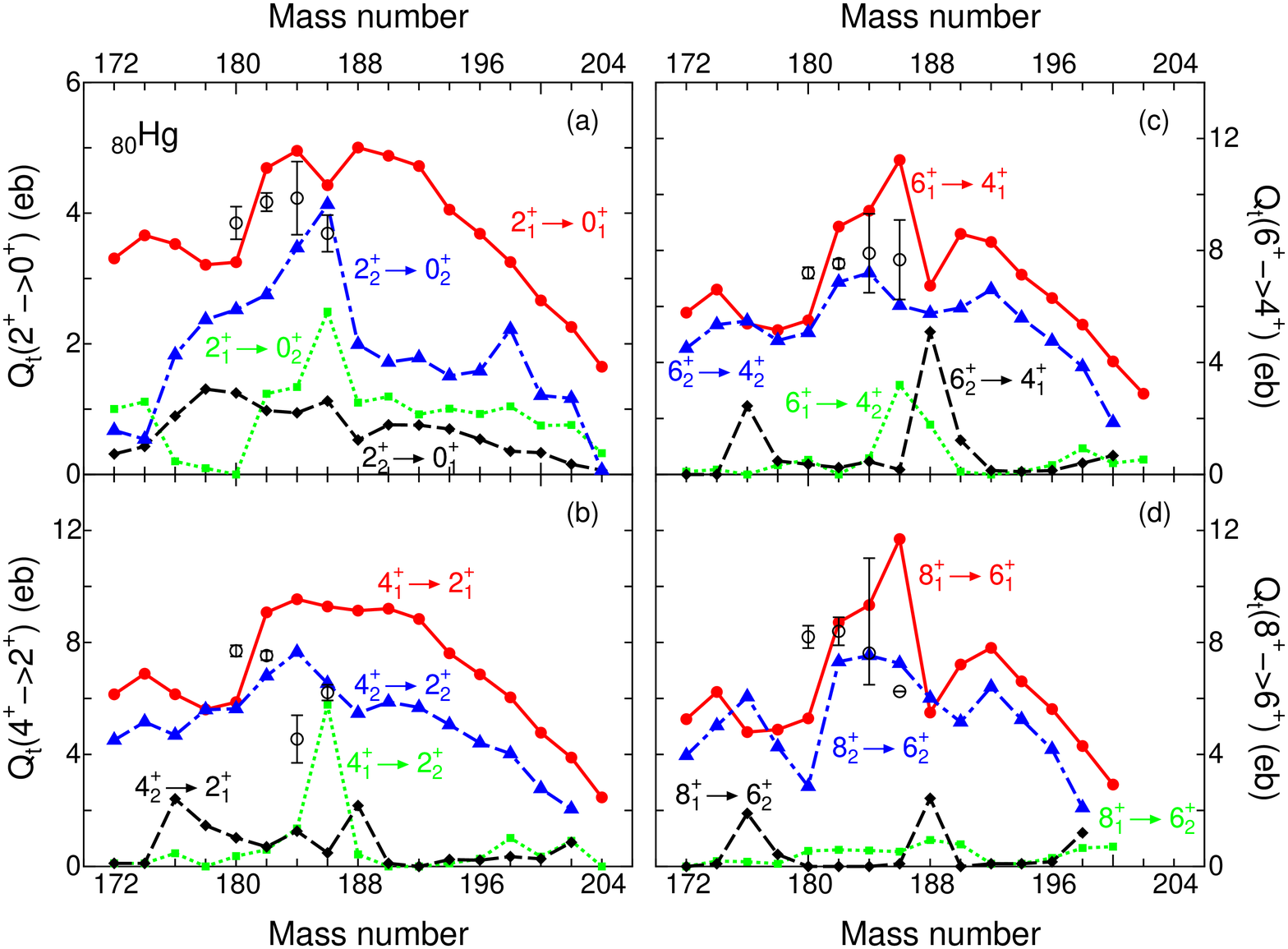}
\caption{(Color online) The calculated transition quadrupole moment 
 $Q_{t}(J\rightarrow J-2)$ for $J^{\pi}=2^{+}$ (a), $4^{+}$ (b),
 $6^{+}$ (c) and $8^{+}$ (d) states compared with the experimental
$Q_{t}$ values for transitions between the yrast states 
 \cite{grahn09,proetel74,ma86} (open circles). }
\label{fig:qt}
\end{center}
\end{figure*}

From the $B({\textnormal{E2}};J\rightarrow J^{\prime})$ transition rates, one can extract the transition
quadrupole moment $Q_{t}(J\rightarrow J^{\prime})$ for which there are a number of
available experimental data to compare with. 
$Q_{t}(J\rightarrow J^{\prime})$ is related to the
$B({\textnormal{E2}};J\rightarrow J^{\prime})$ value through
\begin{eqnarray}
B({\textnormal{E2}};J\rightarrow J^{\prime})=\frac{5}{16\pi}(J200|J^{\prime}0)^{2}\{Q_{t}(J\rightarrow J^{\prime})\}^{2}
\end{eqnarray}
Figure \ref{fig:qt} exhibits the calculated $Q_{t}(J\rightarrow J-2)$
value for $J^{\pi}=2^{+}$ (a), $4^{+}$ (b), $6^{+}$ (c) and $8^{+}$ (d) states compared with
the experimental $Q_{t}$ values for the transition between the yrast
states \cite{proetel74,ma86,grahn09}. 
In each panel, the transitions between the yrast $J^{+}$ and $(J-2)^{+}$
states and between the non-yrast ones are strong in most of the nuclei. 
In the nuclei around the mid-shell nucleus $^{184}$Hg, the
transition between the yrast states correspond to the in-band E2 transitions within 
$0p$-$0h$ or $2p$-$2h$ band with strong collectivity and are particularly
large in Fig.~\ref{fig:qt}. 
One of the $Q_{t}$ values for the two in-band transitions follows the experimental data. 
On the other hand, the transitions 
between the yrast $J^{+}$ and the non-yrast $(J-2)^{+}$, or vice versa, are
overall weak other than the $^{186}$Hg nucleus in
Figs.~\ref{fig:qt}(a), \ref{fig:qt}(b) and \ref{fig:qt}(c), and the
$^{188}$Hg in Figs.~\ref{fig:qt}(c) and \ref{fig:qt}(d), where the mixing between
the two configurations turned out to be significant in the present
calculation.

One can also deduce the deformation parameter 
$\beta_{t}(J\rightarrow J-2)$ from $Q_{t}$ through the relation
\begin{eqnarray}
\label{eq:b2}
\beta_{t}(J\rightarrow
J-2)=\frac{\sqrt{5\pi}}{3ZR^{2}}Q_{t}(J\rightarrow J-2), 
\end{eqnarray}
where $R=1.2 A^{1/3}$ fm.  
As examples, the calculated values $\beta_{t}(2^{+}_{1}\rightarrow
0^{+}_{1})\approx 0.15-0.17$ of $^{184,186,188}$Hg,
corresponding to the oblate configuration, is consistent with the
experimental values $\beta_{t}(2^{+}_{1}\rightarrow
0^{+}_{1})=0.15(2)$ for $^{184}$Hg \cite{ma86} and $0.13(1)$ for $^{186}$Hg \cite{proetel74}, 
and as well as with the minimum at $\beta\approx 0.15$ in the
mean-field energy surfaces in Fig.~\ref{fig:pes1}. 
However, the present $\beta_{t}(2^{+}_{2}\rightarrow 0^{+}_{2})$ value 
for the $^{186}$Hg ($^{188}$Hg) nucleus, corresponding to the prolate deformation, is too small,
$\beta_{t}(2^{+}_{2}\rightarrow 0^{+}_{2})\approx 0.15$ (0.17) than the
prolate mean-field minimum at $\beta\approx 0.3$ (cf. Fig.~\ref{fig:pes1}). 
The reason are the too strong mixing in these nuclei and also the $\gamma$-softness in the
prolate configuration. 
For $^{182}$Hg ($^{180}$Hg), where the prolate band is predicted to be the lowest band 
(cf. Fig.~\ref{fig:182}), $\beta_{t}(2^{+}_{1}\rightarrow
0^{+}_{1})=0.12$ (0.12), which is again too small compared to the prolate mean-field
minimum at $\beta\approx 0.3$, whereas the present $\beta_{t}(2^{+}_{2}\rightarrow
0^{+}_{2})=0.12$ (0.09) for oblate configuration agrees with the oblate
mean-field minimum at $\beta\approx 0.15$.


\subsection{Ground-state properties}

\begin{figure}[ctb!]
\begin{center}
\includegraphics[width=8.6cm]{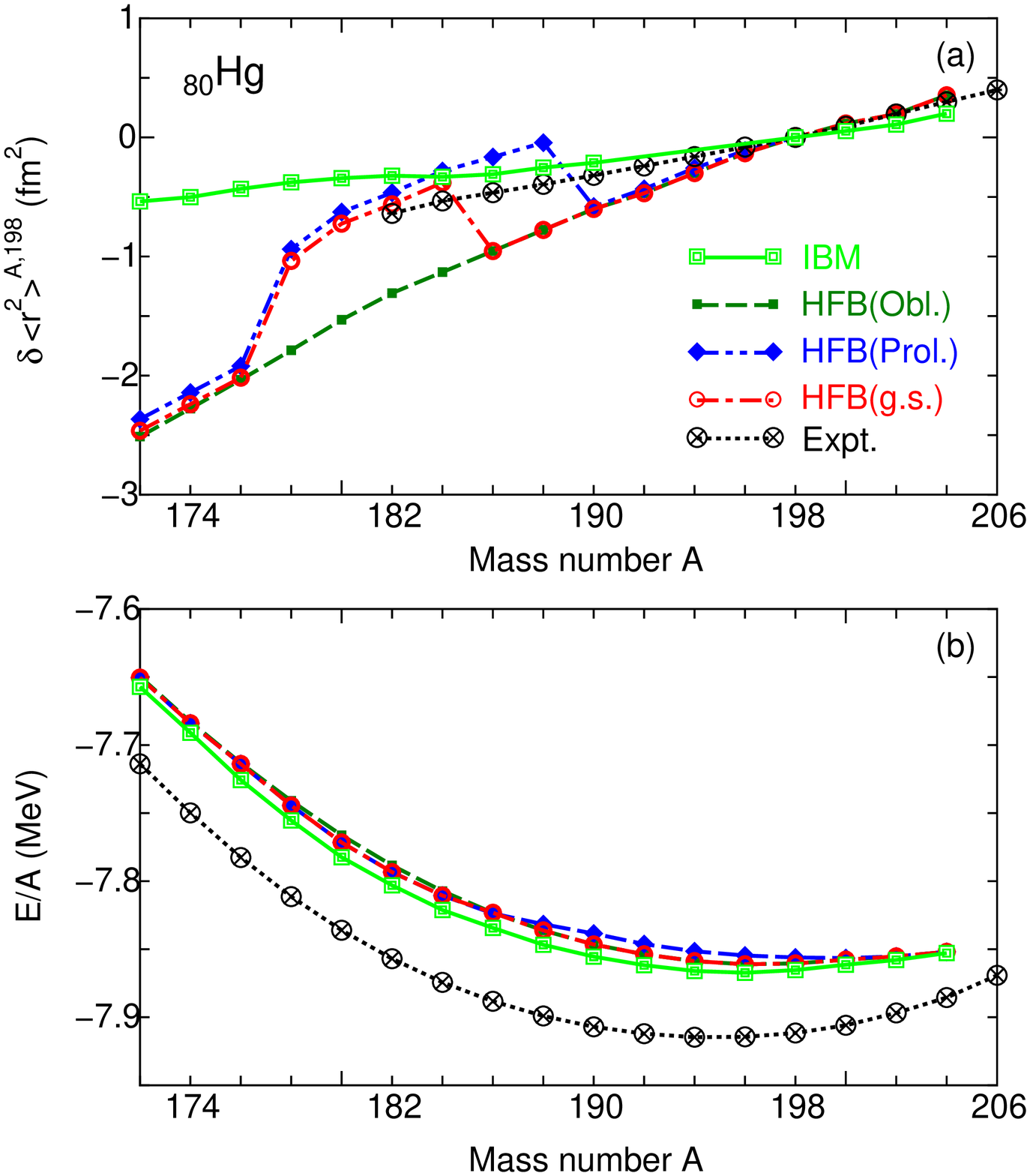}
\caption{(Color online) The mean square charge radii $\delta\langle
 r^{2}\rangle^{A,198}$ (a) and the ground-state energy per nucleon $E/A$
 (b) for the considered $^{172-204}$Hg.  
In each panel, the HFB results corresponding to the oblate and the
 prolate minima, as well as the global minima (one of the former two),
 and the IBM mixing results for the $0^{+}_{1}$ state are shown. 
The experimental data have been taken from \cite{angeli04} and
 \cite{mass_data} for $\delta\langle
 r^{2}\rangle^{A,198}$ and $E/A$, respectively. }
\label{fig:radii}
\end{center}
\end{figure}

It is worthwhile to compare the ground-state properties of
the considered Hg nuclei obtained with the mean field calculation, 
with the wealth of the available experimental data. 
In this section we analyze the mean square charge radii and the binding
energies. The former plays a relevant role as an indicator of the character of 
the ground state deformation. 

In the HFB method, the charge radius is obtained as the mean value of
the $r^{2}$ operator for each of the oblate and prolate
minima. 
In the configuration mixing IBM-2 framework, the charge radius 
$\langle r^{2}\rangle$ is connected to the matrix element of the
E0 operator. 
The E0 operator $\hat T^{\textnormal{(E0)}}$ is given as \cite{IBM}
\begin{eqnarray}
\label{eq:e0}
\hat T^{\textnormal{(E0)}}=\sum_{i=1,3}\sum_{\rho=\nu,\pi}\hat{\cal
 P}_{i}(\beta_{i,\rho}\hat n_{d\rho,i}+\gamma_{i,\rho}
 N_{\rho,i})\hat{\cal P}_{i}, 
\end{eqnarray}
with four parameters $\beta_{i,\rho}$ and $\gamma_{i,\rho}$. 
The mean square radius $\langle r^{2}\rangle$ is written as 
\begin{eqnarray}
 \langle r^{2}\rangle=\langle r^{2}_{c}\rangle+\langle\hat
  T^{\textnormal{(E0)}}\rangle_{0^{+}_{1}}. 
\end{eqnarray}
$\langle r^{2}_{c}\rangle$ represents the contribution from the
inert core, which is omitted here since we discuss the $\langle
r^{2}\rangle$ values relative to a particular nucleus. 
For the parameters in the E0 operator of the IBM-2, we adapt the values 
used in the study of the isomer shift in $^{184-200}$Hg
\cite{Barfield83}, $\beta_{1,\nu}=\beta_{3,\nu}=-0.068$ fm$^{2}$, 
$\beta_{1,\pi}=\beta_{3,\pi}=0$ fm$^{2}$, $\gamma_{1,\nu}+\gamma_{3,\nu}=-0.083$ fm$^{2}$
and $\gamma_{1,\pi}=\gamma_{3,\pi}=0$ fm$^{2}$. 
For $\gamma_{1,\nu}$ and $\gamma_{3,\nu}$, the average
$\gamma_{1,\nu}=\gamma_{3,\nu}=-0.0415$ fm$^{2}$ is taken. 
Also, to make the radius in the IBM-2 change linearly when crossing the mid-shell,
we approximately correct the boson number $N_{\nu,i}$ in 
Eq.~(\ref{eq:e0}) so that 
$N_{\nu}$ runs from 2 ($^{204}$Hg) to 17 ($^{172}$Hg) and should be replaced with 
$N_{\nu}^{\prime}=(126-2N_{\nu})/2$. 

In Fig.~\ref{fig:radii}(a) we compare the mean square charge radii relative
to the $^{198}$Hg nucleus, $\delta\langle r^{2}\rangle^{A,198}=\langle
r^{2}\rangle^{A}-\langle r^{2}\rangle^{198}$, calculated within the HFB
and the IBM-2, with the experimental data taken from \cite{angeli04}. 
The HFB results for both the oblate and prolate minima have similar values
for $A=174-176$ and $A=190-204$ and change linearly with mass. 
But around the near mid-shell nucleus $^{184}$Hg, the HFB charge radius computed
with the
prolate minimum wave function becomes significantly larger than the one for the oblate
minimum, being in a better agreement with the data for these nuclei
exhibiting the prolate intruder configuration. The reason for this behaviour
is the large quadrupole deformation for the minima in these nuclei. In this
region, shape mixing would lead to a ground state charge radius in between
the prolate and oblate results. Both the HFB radii obtained
at the prolate and oblate minima have a linear behaviour with mass number
similar to the experimental data.
For the IBM-2 result, on the other hand, one should also notice the
linear change with $A$, which furthermore turns out to be quite consistent with the
data for $^{182-204}$Hg, apart from a potential ambiguity in a particular
choice of the E0 parameters. 
A future experiment should clarify how the $\delta\langle r^{2}\rangle$
systematics is extrapolated to $A\le 180$. 

Using the E0 operator in Eq.~(\ref{eq:e0}), we also calculate the
$\rho^{2}_{\textnormal{E0}}$ value between $0^{+}_{1}$ and $0^{+}_{2}$
states. 
$\rho_{\textnormal{E0}}(0^{+}_{2}\rightarrow
  0^{+}_{1})$ is written as 
\begin{eqnarray}
 \rho_{\textnormal{E0}}(0^{+}_{2}\rightarrow
  0^{+}_{1})=\frac{Z}{R^{2}}\langle 0^{+}_{1}|\hat
  T^{\textnormal{(E0)}}|0^{+}_{2}\rangle
\end{eqnarray}
To compare with a few available data for $^{184}$Hg ($^{188}$Hg) 
nucleus, the calculated and the experimental \cite{kibedi05} 
$\rho^{2}_{\textnormal{E0}}(0^{+}_{2}\rightarrow 0^{+}_{1})\times 10^{3}$ values are
4.670 (1.447) and 3.2$\pm$1.1 (7$\pm$3), respectively, which are in the
same order of magnitude. 

It is also possible, in the present framework, to compare the calculated
binding energy with the experiment. 
Figure \ref{fig:radii}(b) displays the comparison between theoretical
and the experimental \cite{mass_data} ground-state energy per nucleon $E/A$. 
The HFB results are based on the mean-field ground-state energies for oblate and
prolate minima. 
In the IBM-2, on the other hand, the ground-state energy is obtained by
including the global term in the Hamiltonian Eq.~(\ref{eq:ham-cm}) which
linearly depends on the number of bosons and is irrelevant to the
deformation/excitation \cite{IBM}. 
The global term is determined by adjusting the minimum of the
boson energy surface to the HFB minimum (see \cite{Nom10}, for details). 
We observe in Fig.~\ref{fig:radii}(b) that the calculated $E/A$ for both IBM and HFB ground
states exhibits similar pattern with mass number with respect to
$A\approx 196$ but that suggests a systematic underbinding by $\approx 50$
keV in energy compare to the experimental
data \cite{mass_data}. 
The relativistic Hartree-Bogoliubov calculation of $E/A$ in Hg isotopes with the NL-SC 
functional also suggested \cite{Nik02sc} underbinding but those results 
with a deviation from the experiment $\approx 10$ keV
are more accurate than our result.

\section{Summary\label{sec:summary}}

The method used to derive the parameters of the Hamiltonian of the interacting
boson model with configuration mixing from the constrained HFB calculations with
Gogny D1M energy density functional has been applied to analyze the shape
evolution and the relevant systematics of the low-lying collective states
in Hg isotopes. 
The two independent Hamiltonians corresponding to the $0p$-$0h$ and the
$2p$-$2h$ configurations, and the parameters relevant to the mixing, are
derived without any fit to the data, by mapping the microscopic
constrained energy surface onto the appropriate IBM-2 Hamiltonian in the
boson condensate. 
The energy levels, $B$(E2) transition rates, quadrupole moments, and
some ground-state properties (mean square charge radii and binding energies) are computed from this procedure. 

From the microscopic mean-field calculation (cf. Figs.\ref{fig:pes1} and
\ref{fig:pes2}) we observed: (i) a near spherical 
ground state shape with weak prolate deformation at $\beta\approx 0.08-0.10$ in $^{172,174}$Hg, (ii) onset of
second minimum on the prolate axis at $\beta\approx 0.25$ in $^{176}$Hg,
(iii) transition of the first minimum from axial prolate axis to oblate
axis in $^{178}$Hg, (iv) coexistence of oblate ($\beta\approx 0.15$) and
prolate $\beta\approx 0.3$ minima for
$^{178-190}$Hg, (v) disappearance of the prolate minimum in $^{192}$Hg
and the subsequent weakly oblate deformed structure from $^{192}$Hg to
around $^{198}$Hg, and (vi) near spherical vibrational structure in
$^{200-204}$Hg approaching the neutron $N=126$ shell closure.

The energy levels resulting from the mapped IBM-2 Hamiltonian for $^{172-174,192-204}$Hg
nuclei with a single configuration follows the experimental trend rather
well, and they correlate with the expectations from the microscopic calculation
mentioned above. 
Also for the near mid-shell nucleus $^{184}$Hg, the configuration mixing
calculation reveals that the low-lying $0^{+}_{2}$ arises either from
the intruder $2p$-$2h$ or from the normal $0p$-$0h$ configuration. 
The theoretical prediction for $^{186,188}$Hg, that the oblate band is the ground-state band
and that the intruder prolate band is the second lowest band, turned out
to be consistent with the empirical assignment suggested
experimentally. 

Through the investigation of the detailed level scheme for each
individual nucleus showing manifest shape coexistence we can point out the following discrepancies
between the present calculation and experiment: (i) Particularly in
$^{180,182}$Hg, the $0^{+}_{2}$
energy level is too high compared to the data and,
(ii) contrary to the empirical assignment, the prolate intruder band becomes
the ground-state band in $^{180,182,184}$Hg. 
(iii) Overall, level 
structure and $B$(E2) systematics have not been fully reproduced,
characterized by, e.g., too low $2^{+}_{1}$ energy levels and the
stretching in the energy levels of the higher-spin states in each band. 

We have considered several possibilities to explain these
problems: A peculiar topology of the microscopic energy surface and the
too strong mixing between the two configurations in the IBM. 
The first possibility concerns the above-mentioned problems (i) and (ii),
and is attributed to the property of the currently used density
functional itself. 
This is perhaps most related to how the single-particle spectrum looks like in
these $^{180,182}$Hg nuclei, which should determine the shell gap at $Z=82$
and thus the energy to create $2p$-$2h$ excitation of major importance in
the description of the correct oblate-prolate dominance in the mean
field. 
In this respect, as investigated in \cite{Nik02sc}, it would be of
interest to extend the EDF framework encompassing the complex 
nuclei showing different shapes as considered here. 
Concerning the latter possibility related to the mapping procedure,
one could use a smaller mixing strength and offset in order to describe
the correct level energy spacings and $B$(E2) systematics. 
Although these parameters relevant to the mixing are mainly dependent on 
the topology of the microscopic energy surface, 
the present Hamiltonian in Eq.~(\ref{eq:ham-cm}) might be, therefore,
too simple to reproduce every detail of the experimental low-lying
structure. 
In fact, many of the phenomenological IBM calculations with configuration
mixing considered more interaction terms and parameters in the Hamiltonian. 
However, determining even larger number of these parameters, including
effective charges, from a single mean-field energy surface is apparently not reasonable. 
For this reason, an improved/extended mapping procedure to efficiently extract these
interaction strengths may be worth to study.

\begin{acknowledgments}
Authors would like to thank P. von Brentano, M. Hackstein, J. Jolie, T. Otsuka and
 N. Shimizu for useful discussions. 
Author K. N. acknowledges the support of the JSPS Postdoctoral
 Fellowships for Research Abroad. 
L. M. R. acknowledges support of MINECO through grants Nos. FPA2009-08958 and
FIS2009-07277 as well as the Consolider-Ingenio 2010 program CPAN 
CSD2007-00042 and MULTIDARK  CSD2009-00064.
\end{acknowledgments}

\appendix

\section{Procedure to extract parameters for the IBM Hamiltonian with
 configuration mixing \label{sec:mapping}}

A number of parameters are involved in the configuration mixing IBM
Hamiltonian in Eq.~(\ref{eq:ham-cm}), including offset energy $\Delta_{intr}$ and mixing strength $\omega$. 
It is, therefore, not feasible to determine these parameters
simultaneously through the mapping of the 
microscopic energy surface onto the boson energy surface. 
It is then necessary to determine the parameters with certain approximations. 

First, we fix the parameters for each individual Hamiltonian $\hat
H_{1}$ and $\hat H_{3}$ (cf. Eq.~(\ref{eq:ham-cm})). 
This is done by fitting the coherent-state expectation value of the $0p$-$0h$ 
($E_{11}(\beta,\gamma)$) and
the $2p$-$2h$ ($E_{33}(\beta,\gamma)$) Hamiltonians (cf. Eq.~(\ref{eq:pes-cm})) to the oblate 
$\beta_{min}\approx 0.15$ and to the prolate $\beta_{min}\approx 0.3$
minima, respectively. 
We here assume that the mean-field energy surface can be separated into
two parts in terms of $\gamma$ variable, namely,
$0^{\circ}\le\gamma\le\gamma_{bar}$ and $\gamma_{bar}\le\gamma\le
60^{\circ}$ for prolate and oblate configurations, respectively. 
Here $\gamma_{bar}$ ($0^{\circ}\le\gamma_{bar}\le 60^{\circ}$)
denotes the $\gamma$ value corresponding to the barrier between the two
minima ($\gamma_{bar}=25^{\circ}$ for $^{186}$Hg).  
In the case of $^{186}$Hg, for instance, within the ranges $25^{\circ}\le\gamma\le 60^{\circ}$ and
$0^{\circ}\le\gamma\le 25^{\circ}$ for the $0p$-$0h$ and the $2p$-$2h$
Hamiltonians, respectively, the parameters for $\hat
H_{1}$ and $\hat H_{3}$ can be fixed separately, using the method of 
Refs.~\cite{Nom08,Nom10}: The boson energy surface matches the
microscopic energy surface in the basic topology only in the
neighbourhood of each minimum, i.e., curvatures in both $\beta$ and
$\gamma$ directions up to typically 2 MeV in energy from the minimum, of the
microscopic energy surface. 
In this way, the mean effect of the fermion properties relevant to
determining the low-energy structure of a given nucleus is simulated in
the boson system \cite{Nom08,Nom10}. 
Note that, at this point, the mixing 
interaction is not introduced yet.

Secondly, having determined the strength parameters for each individual
unperturbed Hamiltonian $\hat H_{i}$ ($i=1,3$) $\epsilon_{i}^{\prime}$,
$\kappa_{i}$, $\chi_{\nu,i}$, $\chi_{\pi,i}$, 
$\kappa_{i}^{\prime\prime}$ and $C_{\beta,i}$, which also appeared in 
Eq.~(\ref{eq:pes-diag}), we then obtain the off-set energy
$\Delta_{intr}$ so that the relative energy location of the
oblate and the prolate HFB minima, denoted as $\delta E_{\textnormal{HFB}}^{ob-pr}$, is reproduced:
\begin{eqnarray}
\label{eq:delta}
\delta
E_{\textnormal{HFB}}^{ob-pr}=\{E_{33}(q_{min}^{intr})+\Delta_{intr}\}-E_{11}(q_{min}^{nor}) 
\end{eqnarray}
with $q_{min}^{nor}=(\beta_{min}^{nor},\gamma_{min}^{nor})$ and
$q_{min}^{intr}=(\beta_{min}^{intr},\gamma_{min}^{intr})$ corresponding
to the energy minima for the normal and the intruder configurations,
respectively. 

Finally, the mixing interaction $\hat H_{mix}$ is considered with the
simplification of $\omega_{s}=\omega_{d}=\omega$ (see, main text). 
Partly due to this simplification, the analytical expression of the expectation value of the mixing
interaction has not have an enough flexibility able to reproduce every
detail of the topology around the barrier between the two mean-field
minima. 
Because of this restriction, we assume that the interaction $\hat H_{mix}$ should only
perturbatively contribute to the energy surface, and the $\omega$ value is
fixed so that the overall topology around the barrier becomes similar to
the one in the mean-field energy surface. 
This assumption, as well as the approximate equality in Eq.~(\ref{eq:delta}), seems to be valid, so long as the moderate value
$\omega\approx 0.15-0.22$ MeV, 
which is not too far from value used in the earlier IBM-2
phenomenology on the Hg isotopes \cite{Barfield83,duval81,duval82}, is chosen. 
We have also confirmed that, e.g., in $^{186}$Hg, the oblate and the
prolate minima of the boson energy surface 
changes only by 20-30 keV in energy when the mixing interaction 
$\hat H_{mix}$ is introduced. 

We here comment on the uniqueness of the parameters used in the present
work. 
There may exist other parameter sets which are very different from the one used
here but which equally give a good fit to the microscopic energy surface. 
It is then necessary to adapt one set of parameters, which fits the 
microscopic energy surface but at the same time physically makes sense. 
We consider the following criteria, concerning the range and the
boson-number dependencies, of the parameters so as to be more or less consistent with the knowledge from our previous 
results \cite{Nom08,Nom10,Nom12tri} for other isotopic chain and from earlier microscopic study of IBM
base on the shell-model configuration (for instance, \cite{OAI}): (i) $d$-boson energy $\epsilon_{i}$ should
decrease in an isotopic chain with
the number of neutron bosons toward near midshell. (ii)
Quadrupole-quadrupole interaction strength $\kappa_{i}$ should be stable
against nucleon number, but can slightly increase in its magnitude
toward the shell closures. (iii) For oblate (neutron) deformation, the
sign of the sum $\chi_{\nu,i}+\chi_{\pi,i}$ must be positive 
(negative), (iv) $\chi_{\pi,i}$ can change but should be almost constant
with proton boson number $N_{\pi}$, and (v) if the minimum is soft in $\gamma$, the cubic term
should have the non-zero interaction strength with 
the typical range $\kappa^{\prime\prime}_{i}\approx 0.1-0.2$ MeV
according to our
earlier study \cite{Nom12tri} and other IBM-1 phenomenology, and the sum  
$|\chi_{\nu,i}+\chi_{\pi,i}|$ should be small in magnitude. 

By performing the approximate mapping procedure step-by-step, we have
employed the set of parameters which best fits
the microscopic energy surface and which satisfies the conditions consistent
with our earlier results. 
Nevertheless, due to a large number of parameters, restrictions in the
analytical formula of the IBM energy surface and approximations
mentioned above, taking a 
quantitative measure to evaluate the quality of the mapping is not as
simple as in the case of a single configuration/minimum.

Finally, we mention the difference between the procedure to determine the
off-set energy $\Delta_{intr}$ in the present paper and the procedure taken in
our previously published work on the Pb isotopes exhibiting spherical,
oblate and prolate mean-field minima \cite{Nom12sc}. 
In Ref.~\cite{Nom12sc}, the offset energy (denoted here as $\Delta_{A}$)
was fixed to reproduce 
the energy difference between the spherical and the oblate/prolate HFB
minima, as shown in Eq.~(5) in \cite{Nom12sc}. 
Thus, the procedure to extract the $\Delta_{A}$ value in
Ref.~\cite{Nom12sc} is exactly the same as the one taken in the present paper to determine
the $\Delta_{intr}$ value through Eq.~(\ref{eq:delta}). 
In Ref.~\cite{Nom12sc}, however, when diagonalizing the full Hamiltonian
the $\Delta_{A}$ value had to be corrected by replacing it with the one defined in terms of the
unperturbed $0^{+}_{1}$ eigenenergies, denoted here as $\Delta_{B}$,
(see Eq.~(8) in Ref.~\cite{Nom12sc} and Appendix C of Ref.~\cite{duval82}). 
The reason was that the $0p$-$0h$ Hamiltonian used in \cite{Nom12sc} did
not gain the correlation energy (the $\hat Q_{\nu}\cdot\hat Q_{\pi}$ interaction
term vanishes for the $0p$-$0p$
spherical configuration with $N_{\pi}=0$ in Pb nuclei) so that the $\Delta_{A}$ was too small 
a value to reproduce the empirical spherical-oblate/prolate band structure. 
In the present work, on the other side, since the $0p$-$0h$ configuration gains correlation 
energy for the $^{176-190}$Hg isotopes considered here, we do not
need to correct the $\Delta_{intr}$ value and use it in diagonalization without any modification. 
This is in contrast with the replacement of $\Delta_{A}$ with $\Delta_{B}$
in Ref.~\cite{Nom12sc}.

\bibliography{refs}

\begin{thebibliography}{79}
\expandafter\ifx\csname natexlab\endcsname\relax\def\natexlab#1{#1}\fi
\expandafter\ifx\csname bibnamefont\endcsname\relax
  \def\bibnamefont#1{#1}\fi
\expandafter\ifx\csname bibfnamefont\endcsname\relax
  \def\bibfnamefont#1{#1}\fi
\expandafter\ifx\csname citenamefont\endcsname\relax
  \def\citenamefont#1{#1}\fi
\expandafter\ifx\csname url\endcsname\relax
  \def\url#1{\texttt{#1}}\fi
\expandafter\ifx\csname urlprefix\endcsname\relax\def\urlprefix{URL }\fi
\providecommand{\bibinfo}[2]{#2}
\providecommand{\eprint}[2][]{\url{#2}}

\bibitem[{\citenamefont{Heyde et~al.}(1983)\citenamefont{Heyde, Van~Isacker,
  Waroquier, Wood, and Meyer}}]{heyde83}
\bibinfo{author}{\bibfnamefont{K.}~\bibnamefont{Heyde}},
  \bibinfo{author}{\bibfnamefont{P.}~\bibnamefont{Van~Isacker}},
  \bibinfo{author}{\bibfnamefont{M.}~\bibnamefont{Waroquier}},
  \bibinfo{author}{\bibfnamefont{J.~L.} \bibnamefont{Wood}}, \bibnamefont{and}
  \bibinfo{author}{\bibfnamefont{R.~A.} \bibnamefont{Meyer}},
  \bibinfo{journal}{Phys. Rep.} \textbf{\bibinfo{volume}{102}},
  \bibinfo{pages}{291 } (\bibinfo{year}{1983}).

\bibitem[{\citenamefont{Wood et~al.}(1992)\citenamefont{Wood, Heyde,
  Nazarewicz, Huyse, and van Duppen}}]{wood92}
\bibinfo{author}{\bibfnamefont{J.~L.} \bibnamefont{Wood}},
  \bibinfo{author}{\bibfnamefont{K.}~\bibnamefont{Heyde}},
  \bibinfo{author}{\bibfnamefont{W.}~\bibnamefont{Nazarewicz}},
  \bibinfo{author}{\bibfnamefont{M.}~\bibnamefont{Huyse}}, \bibnamefont{and}
  \bibinfo{author}{\bibfnamefont{P.}~\bibnamefont{van Duppen}},
  \bibinfo{journal}{Phys. Rep.} \textbf{\bibinfo{volume}{215}},
  \bibinfo{pages}{101 } (\bibinfo{year}{1992}).

\bibitem[{\citenamefont{Andreyev et~al.}(2000)\citenamefont{Andreyev, Huyse,
  Van~Duppen, Weissman, Ackermann, Gerl, Hessberger, Hofmann, Kleinb\"ohl,
  M\"unzenberg et~al.}}]{andre00}
\bibinfo{author}{\bibfnamefont{A.~N.} \bibnamefont{Andreyev}},
  \bibinfo{author}{\bibfnamefont{M.}~\bibnamefont{Huyse}},
  \bibinfo{author}{\bibfnamefont{P.}~\bibnamefont{Van~Duppen}},
  \bibinfo{author}{\bibfnamefont{L.}~\bibnamefont{Weissman}},
  \bibinfo{author}{\bibfnamefont{D.}~\bibnamefont{Ackermann}},
  \bibinfo{author}{\bibfnamefont{J.}~\bibnamefont{Gerl}},
  \bibinfo{author}{\bibfnamefont{F.~P.} \bibnamefont{Hessberger}},
  \bibinfo{author}{\bibfnamefont{S.}~\bibnamefont{Hofmann}},
  \bibinfo{author}{\bibfnamefont{A.}~\bibnamefont{Kleinb\"ohl}},
  \bibinfo{author}{\bibfnamefont{G.}~\bibnamefont{M\"unzenberg}},
  \bibnamefont{et~al.}, \bibinfo{journal}{Nature (London)}
  \textbf{\bibinfo{volume}{405}}, \bibinfo{pages}{430} (\bibinfo{year}{2000}).

\bibitem[{\citenamefont{Julin et~al.}(2001)\citenamefont{Julin, Helariutta, and
  Muikku}}]{julin01rev}
\bibinfo{author}{\bibfnamefont{R.}~\bibnamefont{Julin}},
  \bibinfo{author}{\bibfnamefont{K.}~\bibnamefont{Helariutta}},
  \bibnamefont{and} \bibinfo{author}{\bibfnamefont{M.}~\bibnamefont{Muikku}},
  \bibinfo{journal}{J. Phys. G} \textbf{\bibinfo{volume}{27}},
  \bibinfo{pages}{R109} (\bibinfo{year}{2001}).

\bibitem[{\citenamefont{Heyde and Wood}(2011)}]{heyde11}
\bibinfo{author}{\bibfnamefont{K.}~\bibnamefont{Heyde}} \bibnamefont{and}
  \bibinfo{author}{\bibfnamefont{J.~L.} \bibnamefont{Wood}},
  \bibinfo{journal}{Rev. Mod. Phys.} \textbf{\bibinfo{volume}{83}},
  \bibinfo{pages}{1467} (\bibinfo{year}{2011}).

\bibitem[{\citenamefont{Federman and Pittel}(1977)}]{federman77}
\bibinfo{author}{\bibfnamefont{P.}~\bibnamefont{Federman}} \bibnamefont{and}
  \bibinfo{author}{\bibfnamefont{S.}~\bibnamefont{Pittel}},
  \bibinfo{journal}{Phys. Lett. B} \textbf{\bibinfo{volume}{69}},
  \bibinfo{pages}{385 } (\bibinfo{year}{1977}).

\bibitem[{\citenamefont{Van~Duppen et~al.}(1984)\citenamefont{Van~Duppen,
  Coenen, Deneffe, Huyse, Heyde, and Van~Isacker}}]{duppen84}
\bibinfo{author}{\bibfnamefont{P.}~\bibnamefont{Van~Duppen}},
  \bibinfo{author}{\bibfnamefont{E.}~\bibnamefont{Coenen}},
  \bibinfo{author}{\bibfnamefont{K.}~\bibnamefont{Deneffe}},
  \bibinfo{author}{\bibfnamefont{M.}~\bibnamefont{Huyse}},
  \bibinfo{author}{\bibfnamefont{K.}~\bibnamefont{Heyde}}, \bibnamefont{and}
  \bibinfo{author}{\bibfnamefont{P.}~\bibnamefont{Van~Isacker}},
  \bibinfo{journal}{Phys. Rev. Lett.} \textbf{\bibinfo{volume}{52}},
  \bibinfo{pages}{1974} (\bibinfo{year}{1984}).

\bibitem[{\citenamefont{Heyde et~al.}(1985)\citenamefont{Heyde, Van~Isacker,
  Casten, and Wood}}]{heyde85}
\bibinfo{author}{\bibfnamefont{K.}~\bibnamefont{Heyde}},
  \bibinfo{author}{\bibfnamefont{P.}~\bibnamefont{Van~Isacker}},
  \bibinfo{author}{\bibfnamefont{R.~F.} \bibnamefont{Casten}},
  \bibnamefont{and} \bibinfo{author}{\bibfnamefont{J.~L.} \bibnamefont{Wood}},
  \bibinfo{journal}{Phys. Lett. B} \textbf{\bibinfo{volume}{155}},
  \bibinfo{pages}{303 } (\bibinfo{year}{1985}).

\bibitem[{\citenamefont{Heyde et~al.}(1987)\citenamefont{Heyde, Jolie, Moreau,
  Ryckebusch, Waroquier, Van~Duppen, Huyse, and Wood}}]{heyde87}
\bibinfo{author}{\bibfnamefont{K.}~\bibnamefont{Heyde}},
  \bibinfo{author}{\bibfnamefont{J.}~\bibnamefont{Jolie}},
  \bibinfo{author}{\bibfnamefont{J.}~\bibnamefont{Moreau}},
  \bibinfo{author}{\bibfnamefont{J.}~\bibnamefont{Ryckebusch}},
  \bibinfo{author}{\bibfnamefont{M.}~\bibnamefont{Waroquier}},
  \bibinfo{author}{\bibfnamefont{P.}~\bibnamefont{Van~Duppen}},
  \bibinfo{author}{\bibfnamefont{M.}~\bibnamefont{Huyse}}, \bibnamefont{and}
  \bibinfo{author}{\bibfnamefont{J.~L.} \bibnamefont{Wood}},
  \bibinfo{journal}{Nucl. Phys. A} \textbf{\bibinfo{volume}{466}},
  \bibinfo{pages}{189 } (\bibinfo{year}{1987}).

\bibitem[{\citenamefont{Heyde and Meyer}(1988)}]{heyde88}
\bibinfo{author}{\bibfnamefont{K.}~\bibnamefont{Heyde}} \bibnamefont{and}
  \bibinfo{author}{\bibfnamefont{R.~A.} \bibnamefont{Meyer}},
  \bibinfo{journal}{Phys. Rev. C} \textbf{\bibinfo{volume}{37}},
  \bibinfo{pages}{2170} (\bibinfo{year}{1988}).

\bibitem[{\citenamefont{Heyde et~al.}(1995)\citenamefont{Heyde, Jolie, Lehmann,
  De~Coster, and Wood}}]{heyde95}
\bibinfo{author}{\bibfnamefont{K.}~\bibnamefont{Heyde}},
  \bibinfo{author}{\bibfnamefont{J.}~\bibnamefont{Jolie}},
  \bibinfo{author}{\bibfnamefont{H.}~\bibnamefont{Lehmann}},
  \bibinfo{author}{\bibfnamefont{C.}~\bibnamefont{De~Coster}},
  \bibnamefont{and} \bibinfo{author}{\bibfnamefont{J.~L.} \bibnamefont{Wood}},
  \bibinfo{journal}{Nucl. Phys. A} \textbf{\bibinfo{volume}{586}},
  \bibinfo{pages}{1 } (\bibinfo{year}{1995}).

\bibitem[{\citenamefont{Elseviers et~al.}(2011)\citenamefont{Elseviers,
  Andreyev, Antalic, Barzakh, Bree, Cocolios, Comas, Diriken, Fedorov,
  Fedosseyev et~al.}}]{elseviers11}
\bibinfo{author}{\bibfnamefont{J.}~\bibnamefont{Elseviers}},
  \bibinfo{author}{\bibfnamefont{A.~N.} \bibnamefont{Andreyev}},
  \bibinfo{author}{\bibfnamefont{S.}~\bibnamefont{Antalic}},
  \bibinfo{author}{\bibfnamefont{A.}~\bibnamefont{Barzakh}},
  \bibinfo{author}{\bibfnamefont{N.}~\bibnamefont{Bree}},
  \bibinfo{author}{\bibfnamefont{T.~E.} \bibnamefont{Cocolios}},
  \bibinfo{author}{\bibfnamefont{V.~F.} \bibnamefont{Comas}},
  \bibinfo{author}{\bibfnamefont{J.}~\bibnamefont{Diriken}},
  \bibinfo{author}{\bibfnamefont{D.}~\bibnamefont{Fedorov}},
  \bibinfo{author}{\bibfnamefont{V.~N.} \bibnamefont{Fedosseyev}},
  \bibnamefont{et~al.}, \bibinfo{journal}{Phys. Rev. C}
  \textbf{\bibinfo{volume}{84}}, \bibinfo{pages}{034307}
  (\bibinfo{year}{2011}).

\bibitem[{\citenamefont{{Brookhaven National Nuclear Data Center}}()}]{data}
\bibinfo{author}{\bibnamefont{{Brookhaven National Nuclear Data Center}}},
  \bibinfo{howpublished}{\url{http://www.nndc.bnl.gov}}.

\bibitem[{\citenamefont{Iachello and Arima}(1987)}]{IBM}
\bibinfo{author}{\bibfnamefont{F.}~\bibnamefont{Iachello}} \bibnamefont{and}
  \bibinfo{author}{\bibfnamefont{A.}~\bibnamefont{Arima}},
  \emph{\bibinfo{title}{The interacting boson model}}
  (\bibinfo{publisher}{Cambridge University Press, Cambridge},
  \bibinfo{year}{1987}).

\bibitem[{\citenamefont{Arima et~al.}(1977)\citenamefont{Arima, Otsuka,
  Iachello, and Talmi}}]{Arima77}
\bibinfo{author}{\bibfnamefont{A.}~\bibnamefont{Arima}},
  \bibinfo{author}{\bibfnamefont{T.}~\bibnamefont{Otsuka}},
  \bibinfo{author}{\bibfnamefont{F.}~\bibnamefont{Iachello}}, \bibnamefont{and}
  \bibinfo{author}{\bibfnamefont{I.}~\bibnamefont{Talmi}},
  \bibinfo{journal}{Phys. Lett. B} \textbf{\bibinfo{volume}{66}},
  \bibinfo{pages}{205 } (\bibinfo{year}{1977}).

\bibitem[{\citenamefont{Otsuka et~al.}(1978{\natexlab{a}})\citenamefont{Otsuka,
  Arima, Iachello, and Talmi}}]{OAIT}
\bibinfo{author}{\bibfnamefont{T.}~\bibnamefont{Otsuka}},
  \bibinfo{author}{\bibfnamefont{A.}~\bibnamefont{Arima}},
  \bibinfo{author}{\bibfnamefont{F.}~\bibnamefont{Iachello}}, \bibnamefont{and}
  \bibinfo{author}{\bibfnamefont{I.}~\bibnamefont{Talmi}},
  \bibinfo{journal}{Phys. Lett. B} \textbf{\bibinfo{volume}{76}},
  \bibinfo{pages}{139 } (\bibinfo{year}{1978}{\natexlab{a}}).

\bibitem[{\citenamefont{Otsuka et~al.}(1978{\natexlab{b}})\citenamefont{Otsuka,
  Arima, and Iachello}}]{OAI}
\bibinfo{author}{\bibfnamefont{T.}~\bibnamefont{Otsuka}},
  \bibinfo{author}{\bibfnamefont{A.}~\bibnamefont{Arima}}, \bibnamefont{and}
  \bibinfo{author}{\bibfnamefont{F.}~\bibnamefont{Iachello}},
  \bibinfo{journal}{Nucl. Phys. A} \textbf{\bibinfo{volume}{309}},
  \bibinfo{pages}{1} (\bibinfo{year}{1978}{\natexlab{b}}).

\bibitem[{\citenamefont{Duval and Barrett}(1981)}]{duval81}
\bibinfo{author}{\bibfnamefont{P.~D.} \bibnamefont{Duval}} \bibnamefont{and}
  \bibinfo{author}{\bibfnamefont{B.~R.} \bibnamefont{Barrett}},
  \bibinfo{journal}{Phys. Lett. B} \textbf{\bibinfo{volume}{100}},
  \bibinfo{pages}{223} (\bibinfo{year}{1981}).

\bibitem[{\citenamefont{Duval and Barrett}(1982)}]{duval82}
\bibinfo{author}{\bibfnamefont{P.~D.} \bibnamefont{Duval}} \bibnamefont{and}
  \bibinfo{author}{\bibfnamefont{B.~R.} \bibnamefont{Barrett}},
  \bibinfo{journal}{Nucl. Phys. A} \textbf{\bibinfo{volume}{376}},
  \bibinfo{pages}{213 } (\bibinfo{year}{1982}).

\bibitem[{\citenamefont{Barfield et~al.}(1983)\citenamefont{Barfield, Barrett,
  Sage, and Duval}}]{Barfield83}
\bibinfo{author}{\bibfnamefont{A.~F.} \bibnamefont{Barfield}},
  \bibinfo{author}{\bibfnamefont{B.~R.} \bibnamefont{Barrett}},
  \bibinfo{author}{\bibfnamefont{K.~A.} \bibnamefont{Sage}}, \bibnamefont{and}
  \bibinfo{author}{\bibfnamefont{P.~D.} \bibnamefont{Duval}},
  \bibinfo{journal}{Z. Phys. A} \textbf{\bibinfo{volume}{311}},
  \bibinfo{pages}{205} (\bibinfo{year}{1983}).

\bibitem[{\citenamefont{De~Coster et~al.}(1999)\citenamefont{De~Coster,
  Decroix, Heyde, Jolie, Lehmann, and Wood}}]{decoster99}
\bibinfo{author}{\bibfnamefont{C.}~\bibnamefont{De~Coster}},
  \bibinfo{author}{\bibfnamefont{B.}~\bibnamefont{Decroix}},
  \bibinfo{author}{\bibfnamefont{K.}~\bibnamefont{Heyde}},
  \bibinfo{author}{\bibfnamefont{J.}~\bibnamefont{Jolie}},
  \bibinfo{author}{\bibfnamefont{H.}~\bibnamefont{Lehmann}}, \bibnamefont{and}
  \bibinfo{author}{\bibfnamefont{J.~L.} \bibnamefont{Wood}},
  \bibinfo{journal}{Nucl. Phys. A} \textbf{\bibinfo{volume}{651}},
  \bibinfo{pages}{31 } (\bibinfo{year}{1999}).

\bibitem[{\citenamefont{Fossion et~al.}(2003)\citenamefont{Fossion, Heyde,
  Thiamova, and Van~Isacker}}]{Fossion03}
\bibinfo{author}{\bibfnamefont{R.}~\bibnamefont{Fossion}},
  \bibinfo{author}{\bibfnamefont{K.}~\bibnamefont{Heyde}},
  \bibinfo{author}{\bibfnamefont{G.}~\bibnamefont{Thiamova}}, \bibnamefont{and}
  \bibinfo{author}{\bibfnamefont{P.}~\bibnamefont{Van~Isacker}},
  \bibinfo{journal}{Phys. Rev. C} \textbf{\bibinfo{volume}{67}},
  \bibinfo{pages}{024306} (\bibinfo{year}{2003}).

\bibitem[{\citenamefont{Garc\'ia-Ramos
  et~al.}(2011)\citenamefont{Garc\'ia-Ramos, Hellemans, and Heyde}}]{Garcia11}
\bibinfo{author}{\bibfnamefont{J.~E.} \bibnamefont{Garc\'ia-Ramos}},
  \bibinfo{author}{\bibfnamefont{V.}~\bibnamefont{Hellemans}},
  \bibnamefont{and} \bibinfo{author}{\bibfnamefont{K.}~\bibnamefont{Heyde}},
  \bibinfo{journal}{Phys. Rev. C} \textbf{\bibinfo{volume}{84}},
  \bibinfo{pages}{014331} (\bibinfo{year}{2011}).

\bibitem[{\citenamefont{Frank et~al.}(2004)\citenamefont{Frank, Van~Isacker,
  and Vargas}}]{frank04}
\bibinfo{author}{\bibfnamefont{A.}~\bibnamefont{Frank}},
  \bibinfo{author}{\bibfnamefont{P.}~\bibnamefont{Van~Isacker}},
  \bibnamefont{and} \bibinfo{author}{\bibfnamefont{C.~E.}
  \bibnamefont{Vargas}}, \bibinfo{journal}{Phys. Rev. C}
  \textbf{\bibinfo{volume}{69}}, \bibinfo{pages}{034323}
  (\bibinfo{year}{2004}).

\bibitem[{\citenamefont{Frank et~al.}(2006)\citenamefont{Frank, Van~Isacker,
  and Iachello}}]{frank06}
\bibinfo{author}{\bibfnamefont{A.}~\bibnamefont{Frank}},
  \bibinfo{author}{\bibfnamefont{P.}~\bibnamefont{Van~Isacker}},
  \bibnamefont{and} \bibinfo{author}{\bibfnamefont{F.}~\bibnamefont{Iachello}},
  \bibinfo{journal}{Phys. Rev. C} \textbf{\bibinfo{volume}{73}},
  \bibinfo{pages}{061302} (\bibinfo{year}{2006}).

\bibitem[{\citenamefont{Irving~Morales
  et~al.}(2008)\citenamefont{Irving~Morales, Frank, Vargas, and
  Van~Isacker}}]{morales08}
\bibinfo{author}{\bibfnamefont{O.}~\bibnamefont{Irving~Morales}},
  \bibinfo{author}{\bibfnamefont{A.}~\bibnamefont{Frank}},
  \bibinfo{author}{\bibfnamefont{C.~E.} \bibnamefont{Vargas}},
  \bibnamefont{and}
  \bibinfo{author}{\bibfnamefont{P.}~\bibnamefont{Van~Isacker}},
  \bibinfo{journal}{Phys. Rev. C} \textbf{\bibinfo{volume}{78}},
  \bibinfo{pages}{024303} (\bibinfo{year}{2008}).

\bibitem[{\citenamefont{Heyde et~al.}(1992)\citenamefont{Heyde, De~Coster,
  Jolie, and Wood}}]{heyde92}
\bibinfo{author}{\bibfnamefont{K.}~\bibnamefont{Heyde}},
  \bibinfo{author}{\bibfnamefont{C.}~\bibnamefont{De~Coster}},
  \bibinfo{author}{\bibfnamefont{J.}~\bibnamefont{Jolie}}, \bibnamefont{and}
  \bibinfo{author}{\bibfnamefont{J.~L.} \bibnamefont{Wood}},
  \bibinfo{journal}{Phys. Rev. C} \textbf{\bibinfo{volume}{46}},
  \bibinfo{pages}{541} (\bibinfo{year}{1992}).

\bibitem[{\citenamefont{De~Coster et~al.}(1996)\citenamefont{De~Coster, Heyde,
  Decroix, Van~Isacker, Jolie, Lehmann, and Wood}}]{DeCoster96}
\bibinfo{author}{\bibfnamefont{C.}~\bibnamefont{De~Coster}},
  \bibinfo{author}{\bibfnamefont{K.}~\bibnamefont{Heyde}},
  \bibinfo{author}{\bibfnamefont{B.}~\bibnamefont{Decroix}},
  \bibinfo{author}{\bibfnamefont{P.}~\bibnamefont{Van~Isacker}},
  \bibinfo{author}{\bibfnamefont{J.}~\bibnamefont{Jolie}},
  \bibinfo{author}{\bibfnamefont{H.}~\bibnamefont{Lehmann}}, \bibnamefont{and}
  \bibinfo{author}{\bibfnamefont{J.~L.} \bibnamefont{Wood}},
  \bibinfo{journal}{Nucl. Phys. A} \textbf{\bibinfo{volume}{600}},
  \bibinfo{pages}{251} (\bibinfo{year}{1996}).

\bibitem[{\citenamefont{Lehmann et~al.}(1997)\citenamefont{Lehmann, Jolie,
  De~Coster, Decroix, Heyde, and Wood}}]{Lehmann97}
\bibinfo{author}{\bibfnamefont{H.}~\bibnamefont{Lehmann}},
  \bibinfo{author}{\bibfnamefont{J.}~\bibnamefont{Jolie}},
  \bibinfo{author}{\bibfnamefont{C.}~\bibnamefont{De~Coster}},
  \bibinfo{author}{\bibfnamefont{B.}~\bibnamefont{Decroix}},
  \bibinfo{author}{\bibfnamefont{K.}~\bibnamefont{Heyde}}, \bibnamefont{and}
  \bibinfo{author}{\bibfnamefont{J.~L.} \bibnamefont{Wood}},
  \bibinfo{journal}{Nucl. Phys. A} \textbf{\bibinfo{volume}{621}},
  \bibinfo{pages}{767 } (\bibinfo{year}{1997}).

\bibitem[{\citenamefont{Bender et~al.}(2003)\citenamefont{Bender, Heenen, and
  Reinhard}}]{Ben03rev}
\bibinfo{author}{\bibfnamefont{M.}~\bibnamefont{Bender}},
  \bibinfo{author}{\bibfnamefont{P.-H.} \bibnamefont{Heenen}},
  \bibnamefont{and} \bibinfo{author}{\bibfnamefont{P.-G.}
  \bibnamefont{Reinhard}}, \bibinfo{journal}{Rev. Mod. Phys.}
  \textbf{\bibinfo{volume}{75}}, \bibinfo{pages}{121} (\bibinfo{year}{2003}).

\bibitem[{\citenamefont{Skyrme}(1958)}]{Skyrme}
\bibinfo{author}{\bibfnamefont{T.~H.~R.} \bibnamefont{Skyrme}},
  \bibinfo{journal}{Nucl. Phys.} \textbf{\bibinfo{volume}{9}},
  \bibinfo{pages}{615} (\bibinfo{year}{1958}).

\bibitem[{\citenamefont{Vautherin and Brink}(1972)}]{VB}
\bibinfo{author}{\bibfnamefont{D.}~\bibnamefont{Vautherin}} \bibnamefont{and}
  \bibinfo{author}{\bibfnamefont{D.~M.} \bibnamefont{Brink}},
  \bibinfo{journal}{Phys. Rev. C} \textbf{\bibinfo{volume}{5}},
  \bibinfo{pages}{626} (\bibinfo{year}{1972}).

\bibitem[{\citenamefont{{J. Decharge and M. Girod and D. Gogny}}(1975)}]{Gogny}
\bibinfo{author}{\bibnamefont{{J. Decharge and M. Girod and D. Gogny}}},
  \bibinfo{journal}{Phys. Lett. B} \textbf{\bibinfo{volume}{55}},
  \bibinfo{pages}{361} (\bibinfo{year}{1975}).

\bibitem[{\citenamefont{Vretenar et~al.}(2005)\citenamefont{Vretenar,
  Afanasjev, Lalazissis, and Ring}}]{vre05rev}
\bibinfo{author}{\bibfnamefont{D.}~\bibnamefont{Vretenar}},
  \bibinfo{author}{\bibfnamefont{A.}~\bibnamefont{Afanasjev}},
  \bibinfo{author}{\bibfnamefont{G.}~\bibnamefont{Lalazissis}},
  \bibnamefont{and} \bibinfo{author}{\bibfnamefont{P.}~\bibnamefont{Ring}},
  \bibinfo{journal}{Phys. Rep.} \textbf{\bibinfo{volume}{409}},
  \bibinfo{pages}{101 } (\bibinfo{year}{2005}).

\bibitem[{\citenamefont{Nik\ifmmode \check{s}\else
  \v{s}\fi{}i\ifmmode~\acute{c}\else \'{c}\fi{}
  et~al.}(2011)\citenamefont{Nik\ifmmode \check{s}\else
  \v{s}\fi{}i\ifmmode~\acute{c}\else \'{c}\fi{}, Vretenar, and
  Ring}}]{Nik11rev}
\bibinfo{author}{\bibfnamefont{T.}~\bibnamefont{Nik\ifmmode \check{s}\else
  \v{s}\fi{}i\ifmmode~\acute{c}\else \'{c}\fi{}}},
  \bibinfo{author}{\bibfnamefont{D.}~\bibnamefont{Vretenar}}, \bibnamefont{and}
  \bibinfo{author}{\bibfnamefont{P.}~\bibnamefont{Ring}},
  \bibinfo{journal}{Prog. Part. Nucl. Phys.} \textbf{\bibinfo{volume}{66}},
  \bibinfo{pages}{519} (\bibinfo{year}{2011}).

\bibitem[{\citenamefont{Nazarewicz}(1993)}]{Naza93}
\bibinfo{author}{\bibfnamefont{W.}~\bibnamefont{Nazarewicz}},
  \bibinfo{journal}{Phys. Lett. B} \textbf{\bibinfo{volume}{305}},
  \bibinfo{pages}{195 } (\bibinfo{year}{1993}).

\bibitem[{\citenamefont{Duguet et~al.}(2003)\citenamefont{Duguet, Bender,
  Bonche, and Heenen}}]{Duguet03}
\bibinfo{author}{\bibfnamefont{T.}~\bibnamefont{Duguet}},
  \bibinfo{author}{\bibfnamefont{M.}~\bibnamefont{Bender}},
  \bibinfo{author}{\bibfnamefont{P.}~\bibnamefont{Bonche}}, \bibnamefont{and}
  \bibinfo{author}{\bibfnamefont{P.-H.} \bibnamefont{Heenen}},
  \bibinfo{journal}{Phys. Lett. B} \textbf{\bibinfo{volume}{559}},
  \bibinfo{pages}{201 } (\bibinfo{year}{2003}).

\bibitem[{\citenamefont{Bender et~al.}(2004)\citenamefont{Bender, Bonche,
  Duguet, and Heenen}}]{Ben04Pb}
\bibinfo{author}{\bibfnamefont{M.}~\bibnamefont{Bender}},
  \bibinfo{author}{\bibfnamefont{P.}~\bibnamefont{Bonche}},
  \bibinfo{author}{\bibfnamefont{T.}~\bibnamefont{Duguet}}, \bibnamefont{and}
  \bibinfo{author}{\bibfnamefont{P.-H.} \bibnamefont{Heenen}},
  \bibinfo{journal}{Phys. Rev. C} \textbf{\bibinfo{volume}{69}},
  \bibinfo{pages}{064303} (\bibinfo{year}{2004}).

\bibitem[{\citenamefont{Girod and Reinhard}(1982)}]{girod82}
\bibinfo{author}{\bibfnamefont{M.}~\bibnamefont{Girod}} \bibnamefont{and}
  \bibinfo{author}{\bibfnamefont{P.~G.} \bibnamefont{Reinhard}},
  \bibinfo{journal}{Phys. Lett. B} \textbf{\bibinfo{volume}{117}},
  \bibinfo{pages}{1 } (\bibinfo{year}{1982}).

\bibitem[{\citenamefont{Delaroche et~al.}(1994)\citenamefont{Delaroche, Girod,
  Bastin, Deloncle, Hannachi, Libert, Porquet, Bourgeois, Hojman, Kilcher
  et~al.}}]{Delaroche94}
\bibinfo{author}{\bibfnamefont{J.~P.} \bibnamefont{Delaroche}},
  \bibinfo{author}{\bibfnamefont{M.}~\bibnamefont{Girod}},
  \bibinfo{author}{\bibfnamefont{G.}~\bibnamefont{Bastin}},
  \bibinfo{author}{\bibfnamefont{I.}~\bibnamefont{Deloncle}},
  \bibinfo{author}{\bibfnamefont{F.}~\bibnamefont{Hannachi}},
  \bibinfo{author}{\bibfnamefont{J.}~\bibnamefont{Libert}},
  \bibinfo{author}{\bibfnamefont{M.~G.} \bibnamefont{Porquet}},
  \bibinfo{author}{\bibfnamefont{C.}~\bibnamefont{Bourgeois}},
  \bibinfo{author}{\bibfnamefont{D.}~\bibnamefont{Hojman}},
  \bibinfo{author}{\bibfnamefont{P.}~\bibnamefont{Kilcher}},
  \bibnamefont{et~al.}, \bibinfo{journal}{Phys. Rev. C}
  \textbf{\bibinfo{volume}{50}}, \bibinfo{pages}{2332} (\bibinfo{year}{1994}).

\bibitem[{\citenamefont{Chasman et~al.}(2001)\citenamefont{Chasman, Egido, and
  Robledo}}]{Chasman01}
\bibinfo{author}{\bibfnamefont{R.~R.} \bibnamefont{Chasman}},
  \bibinfo{author}{\bibfnamefont{J.~L.} \bibnamefont{Egido}}, \bibnamefont{and}
  \bibinfo{author}{\bibfnamefont{L.~M.} \bibnamefont{Robledo}},
  \bibinfo{journal}{Phys. Lett. B} \textbf{\bibinfo{volume}{513}},
  \bibinfo{pages}{325 } (\bibinfo{year}{2001}).

\bibitem[{\citenamefont{Rodr\'iguez-Guzm\'an
  et~al.}(2004)\citenamefont{Rodr\'iguez-Guzm\'an, Egido, and
  Robledo}}]{Rayner04Pb}
\bibinfo{author}{\bibfnamefont{R.~R.} \bibnamefont{Rodr\'iguez-Guzm\'an}},
  \bibinfo{author}{\bibfnamefont{J.~L.} \bibnamefont{Egido}}, \bibnamefont{and}
  \bibinfo{author}{\bibfnamefont{L.~M.} \bibnamefont{Robledo}},
  \bibinfo{journal}{Phys. Rev. C} \textbf{\bibinfo{volume}{69}},
  \bibinfo{pages}{054319} (\bibinfo{year}{2004}).

\bibitem[{\citenamefont{Egido et~al.}(2004)\citenamefont{Egido, Robledo, and
  Rodr\'iguez-Guzm\'an}}]{egido04}
\bibinfo{author}{\bibfnamefont{J.~L.} \bibnamefont{Egido}},
  \bibinfo{author}{\bibfnamefont{L.~M.} \bibnamefont{Robledo}},
  \bibnamefont{and} \bibinfo{author}{\bibfnamefont{R.~R.}
  \bibnamefont{Rodr\'iguez-Guzm\'an}}, \bibinfo{journal}{Phys. Rev. Lett.}
  \textbf{\bibinfo{volume}{93}}, \bibinfo{pages}{082502}
  (\bibinfo{year}{2004}).

\bibitem[{\citenamefont{Moreno et~al.}(2006)\citenamefont{Moreno, Sarriguren,
  \'Alvarez-Rodr\'iguez, and Moya~de Guerra}}]{moreno06}
\bibinfo{author}{\bibfnamefont{O.}~\bibnamefont{Moreno}},
  \bibinfo{author}{\bibfnamefont{P.}~\bibnamefont{Sarriguren}},
  \bibinfo{author}{\bibfnamefont{R.}~\bibnamefont{\'Alvarez-Rodr\'iguez}},
  \bibnamefont{and} \bibinfo{author}{\bibfnamefont{E.}~\bibnamefont{Moya~de
  Guerra}}, \bibinfo{journal}{Phys. Rev. C} \textbf{\bibinfo{volume}{73}},
  \bibinfo{pages}{054302} (\bibinfo{year}{2006}).

\bibitem[{\citenamefont{Nik\ifmmode \check{s}\else
  \v{s}\fi{}i\ifmmode~\acute{c}\else \'{c}\fi{}
  et~al.}(2002)\citenamefont{Nik\ifmmode \check{s}\else
  \v{s}\fi{}i\ifmmode~\acute{c}\else \'{c}\fi{}, Vretenar, Ring, and
  Lalazissis}}]{Nik02sc}
\bibinfo{author}{\bibfnamefont{T.}~\bibnamefont{Nik\ifmmode \check{s}\else
  \v{s}\fi{}i\ifmmode~\acute{c}\else \'{c}\fi{}}},
  \bibinfo{author}{\bibfnamefont{D.}~\bibnamefont{Vretenar}},
  \bibinfo{author}{\bibfnamefont{P.}~\bibnamefont{Ring}}, \bibnamefont{and}
  \bibinfo{author}{\bibfnamefont{G.~A.} \bibnamefont{Lalazissis}},
  \bibinfo{journal}{Phys. Rev. C} \textbf{\bibinfo{volume}{65}},
  \bibinfo{pages}{054320} (\bibinfo{year}{2002}).

\bibitem[{\citenamefont{Bengtsson and Nazarewicz}(1989)}]{bengtsson89}
\bibinfo{author}{\bibfnamefont{R.}~\bibnamefont{Bengtsson}} \bibnamefont{and}
  \bibinfo{author}{\bibfnamefont{W.}~\bibnamefont{Nazarewicz}},
  \bibinfo{journal}{Z. Phys. A} \textbf{\bibinfo{volume}{334}}
  (\bibinfo{year}{1989}).

\bibitem[{\citenamefont{Yao et~al.}(2013)\citenamefont{Yao, Bender, and
  Heenen}}]{yao13}
\bibinfo{author}{\bibfnamefont{J.~M.} \bibnamefont{Yao}},
  \bibinfo{author}{\bibfnamefont{M.}~\bibnamefont{Bender}}, \bibnamefont{and}
  \bibinfo{author}{\bibfnamefont{P.-H.} \bibnamefont{Heenen}},
  \bibinfo{journal}{Phys. Rev. C} \textbf{\bibinfo{volume}{87}},
  \bibinfo{pages}{034322} (\bibinfo{year}{2013}).

\bibitem[{\citenamefont{Nomura et~al.}(2008)\citenamefont{Nomura, Shimizu, and
  Otsuka}}]{Nom08}
\bibinfo{author}{\bibfnamefont{K.}~\bibnamefont{Nomura}},
  \bibinfo{author}{\bibfnamefont{N.}~\bibnamefont{Shimizu}}, \bibnamefont{and}
  \bibinfo{author}{\bibfnamefont{T.}~\bibnamefont{Otsuka}},
  \bibinfo{journal}{Phys. Rev. Lett.} \textbf{\bibinfo{volume}{101}},
  \bibinfo{pages}{142501} (\bibinfo{year}{2008}).

\bibitem[{\citenamefont{Nomura et~al.}(2012{\natexlab{a}})\citenamefont{Nomura,
  Rodr\'iguez-Guzm\'an, Robledo, and Shimizu}}]{Nom12sc}
\bibinfo{author}{\bibfnamefont{K.}~\bibnamefont{Nomura}},
  \bibinfo{author}{\bibfnamefont{R.}~\bibnamefont{Rodr\'iguez-Guzm\'an}},
  \bibinfo{author}{\bibfnamefont{L.~M.} \bibnamefont{Robledo}},
  \bibnamefont{and} \bibinfo{author}{\bibfnamefont{N.}~\bibnamefont{Shimizu}},
  \bibinfo{journal}{Phys. Rev. C} \textbf{\bibinfo{volume}{86}},
  \bibinfo{pages}{034322} (\bibinfo{year}{2012}{\natexlab{a}}).

\bibitem[{\citenamefont{Goriely et~al.}(2009)\citenamefont{Goriely, Hilaire,
  Girod, and P\'eru}}]{D1M}
\bibinfo{author}{\bibfnamefont{S.}~\bibnamefont{Goriely}},
  \bibinfo{author}{\bibfnamefont{S.}~\bibnamefont{Hilaire}},
  \bibinfo{author}{\bibfnamefont{M.}~\bibnamefont{Girod}}, \bibnamefont{and}
  \bibinfo{author}{\bibfnamefont{S.}~\bibnamefont{P\'eru}},
  \bibinfo{journal}{Phys. Rev. Lett.} \textbf{\bibinfo{volume}{102}},
  \bibinfo{pages}{242501} (\bibinfo{year}{2009}).

\bibitem[{\citenamefont{Rodr\'iguez-Guzm\'an
  et~al.}(2010{\natexlab{a}})\citenamefont{Rodr\'iguez-Guzm\'an, Sarriguren,
  Robledo, and Perez-Martin}}]{rayner10odd-1}
\bibinfo{author}{\bibfnamefont{R.}~\bibnamefont{Rodr\'iguez-Guzm\'an}},
  \bibinfo{author}{\bibfnamefont{P.}~\bibnamefont{Sarriguren}},
  \bibinfo{author}{\bibfnamefont{L.~M.} \bibnamefont{Robledo}},
  \bibnamefont{and}
  \bibinfo{author}{\bibfnamefont{S.}~\bibnamefont{Perez-Martin}},
  \bibinfo{journal}{Phys. Lett. B} \textbf{\bibinfo{volume}{691}},
  \bibinfo{pages}{202 } (\bibinfo{year}{2010}{\natexlab{a}}).

\bibitem[{\citenamefont{Rodr\'iguez-Guzm\'an
  et~al.}(2010{\natexlab{b}})\citenamefont{Rodr\'iguez-Guzm\'an, Sarriguren,
  and Robledo}}]{rayner10odd-2}
\bibinfo{author}{\bibfnamefont{R.}~\bibnamefont{Rodr\'iguez-Guzm\'an}},
  \bibinfo{author}{\bibfnamefont{P.}~\bibnamefont{Sarriguren}},
  \bibnamefont{and} \bibinfo{author}{\bibfnamefont{L.~M.}
  \bibnamefont{Robledo}}, \bibinfo{journal}{Phys. Rev. C}
  \textbf{\bibinfo{volume}{82}}, \bibinfo{pages}{044318}
  (\bibinfo{year}{2010}{\natexlab{b}}).

\bibitem[{\citenamefont{Rodr\'iguez-Guzm\'an
  et~al.}(2010{\natexlab{c}})\citenamefont{Rodr\'iguez-Guzm\'an, Sarriguren,
  and Robledo}}]{rayner10odd-3}
\bibinfo{author}{\bibfnamefont{R.}~\bibnamefont{Rodr\'iguez-Guzm\'an}},
  \bibinfo{author}{\bibfnamefont{P.}~\bibnamefont{Sarriguren}},
  \bibnamefont{and} \bibinfo{author}{\bibfnamefont{L.~M.}
  \bibnamefont{Robledo}}, \bibinfo{journal}{Phys. Rev. C}
  \textbf{\bibinfo{volume}{82}}, \bibinfo{pages}{061302}
  (\bibinfo{year}{2010}{\natexlab{c}}).

\bibitem[{\citenamefont{Berger et~al.}(1984)\citenamefont{Berger, Girod, and
  Gogny}}]{D1S}
\bibinfo{author}{\bibfnamefont{J.~F.} \bibnamefont{Berger}},
  \bibinfo{author}{\bibfnamefont{M.}~\bibnamefont{Girod}}, \bibnamefont{and}
  \bibinfo{author}{\bibfnamefont{D.}~\bibnamefont{Gogny}},
  \bibinfo{journal}{Nucl. Phys. A} \textbf{\bibinfo{volume}{428}},
  \bibinfo{pages}{23 } (\bibinfo{year}{1984}).

\bibitem[{\citenamefont{Bohr and Mottelsson}(1975)}]{BM}
\bibinfo{author}{\bibfnamefont{A.}~\bibnamefont{Bohr}} \bibnamefont{and}
  \bibinfo{author}{\bibfnamefont{B.~M.} \bibnamefont{Mottelsson}},
  \emph{\bibinfo{title}{Nuclear Structure}}, vol.~\bibinfo{volume}{2}
  (\bibinfo{publisher}{Benjamin, New York, USA}, \bibinfo{year}{1975}).

\bibitem[{\citenamefont{Nomura et~al.}(2012{\natexlab{b}})\citenamefont{Nomura,
  Shimizu, Vretenar, Nik\ifmmode \check{s}\else
  \v{s}\fi{}i\ifmmode~\acute{c}\else \'{c}\fi{}, and Otsuka}}]{Nom12tri}
\bibinfo{author}{\bibfnamefont{K.}~\bibnamefont{Nomura}},
  \bibinfo{author}{\bibfnamefont{N.}~\bibnamefont{Shimizu}},
  \bibinfo{author}{\bibfnamefont{D.}~\bibnamefont{Vretenar}},
  \bibinfo{author}{\bibfnamefont{T.}~\bibnamefont{Nik\ifmmode \check{s}\else
  \v{s}\fi{}i\ifmmode~\acute{c}\else \'{c}\fi{}}}, \bibnamefont{and}
  \bibinfo{author}{\bibfnamefont{T.}~\bibnamefont{Otsuka}},
  \bibinfo{journal}{Phys. Rev. Lett.} \textbf{\bibinfo{volume}{108}},
  \bibinfo{pages}{132501} (\bibinfo{year}{2012}{\natexlab{b}}).

\bibitem[{\citenamefont{Nomura et~al.}(2011{\natexlab{a}})\citenamefont{Nomura,
  Otsuka, Shimizu, and Guo}}]{Nom11rot}
\bibinfo{author}{\bibfnamefont{K.}~\bibnamefont{Nomura}},
  \bibinfo{author}{\bibfnamefont{T.}~\bibnamefont{Otsuka}},
  \bibinfo{author}{\bibfnamefont{N.}~\bibnamefont{Shimizu}}, \bibnamefont{and}
  \bibinfo{author}{\bibfnamefont{L.}~\bibnamefont{Guo}},
  \bibinfo{journal}{Phys. Rev. C} \textbf{\bibinfo{volume}{83}},
  \bibinfo{pages}{041302} (\bibinfo{year}{2011}{\natexlab{a}}).

\bibitem[{\citenamefont{Ginocchio and Kirson}(1980)}]{GK}
\bibinfo{author}{\bibfnamefont{J.~N.} \bibnamefont{Ginocchio}}
  \bibnamefont{and} \bibinfo{author}{\bibfnamefont{M.~W.}
  \bibnamefont{Kirson}}, \bibinfo{journal}{Nucl. Phys. A}
  \textbf{\bibinfo{volume}{350}}, \bibinfo{pages}{31} (\bibinfo{year}{1980}).

\bibitem[{\citenamefont{Schaaser and Brink}(1986)}]{Schaaser86}
\bibinfo{author}{\bibfnamefont{H.}~\bibnamefont{Schaaser}} \bibnamefont{and}
  \bibinfo{author}{\bibfnamefont{D.~M.} \bibnamefont{Brink}},
  \bibinfo{journal}{Nucl. Phys. A} \textbf{\bibinfo{volume}{452}},
  \bibinfo{pages}{1 } (\bibinfo{year}{1986}).

\bibitem[{\citenamefont{Thouless and Valatin}(1962)}]{TV}
\bibinfo{author}{\bibfnamefont{D.~J.} \bibnamefont{Thouless}} \bibnamefont{and}
  \bibinfo{author}{\bibfnamefont{J.~G.} \bibnamefont{Valatin}},
  \bibinfo{journal}{Nucl. Phys.} \textbf{\bibinfo{volume}{31}},
  \bibinfo{pages}{211 } (\bibinfo{year}{1962}).

\bibitem[{CEA()}]{CEA}
\bibinfo{howpublished}{\url{http://www-phynu.cea.fr/science_en_ligne/carte_pot%
entiels_microscopiques/carte_potentiel_nucleaire_eng.htm}}.

\bibitem[{\citenamefont{Bender et~al.}(2006)\citenamefont{Bender, Bertsch, and
  Heenen}}]{bender06}
\bibinfo{author}{\bibfnamefont{M.}~\bibnamefont{Bender}},
  \bibinfo{author}{\bibfnamefont{G.~F.} \bibnamefont{Bertsch}},
  \bibnamefont{and} \bibinfo{author}{\bibfnamefont{P.-H.}
  \bibnamefont{Heenen}}, \bibinfo{journal}{Phys. Rev. C}
  \textbf{\bibinfo{volume}{73}}, \bibinfo{pages}{034322}
  (\bibinfo{year}{2006}).

\bibitem[{\citenamefont{Nomura et~al.}(2010)\citenamefont{Nomura, Shimizu, and
  Otsuka}}]{Nom10}
\bibinfo{author}{\bibfnamefont{K.}~\bibnamefont{Nomura}},
  \bibinfo{author}{\bibfnamefont{N.}~\bibnamefont{Shimizu}}, \bibnamefont{and}
  \bibinfo{author}{\bibfnamefont{T.}~\bibnamefont{Otsuka}},
  \bibinfo{journal}{Phys. Rev. C} \textbf{\bibinfo{volume}{81}},
  \bibinfo{pages}{044307} (\bibinfo{year}{2010}).

\bibitem[{\citenamefont{Nomura et~al.}(2011{\natexlab{b}})\citenamefont{Nomura,
  Otsuka, Rodr\'iguez-Guzm\'an, Robledo, and Sarriguren}}]{Nom11sys}
\bibinfo{author}{\bibfnamefont{K.}~\bibnamefont{Nomura}},
  \bibinfo{author}{\bibfnamefont{T.}~\bibnamefont{Otsuka}},
  \bibinfo{author}{\bibfnamefont{R.}~\bibnamefont{Rodr\'iguez-Guzm\'an}},
  \bibinfo{author}{\bibfnamefont{L.~M.} \bibnamefont{Robledo}},
  \bibnamefont{and}
  \bibinfo{author}{\bibfnamefont{P.}~\bibnamefont{Sarriguren}},
  \bibinfo{journal}{Phys. Rev. C} \textbf{\bibinfo{volume}{84}},
  \bibinfo{pages}{054316} (\bibinfo{year}{2011}{\natexlab{b}}).

\bibitem[{\citenamefont{Sandzelius et~al.}(2009)\citenamefont{Sandzelius,
  Ganio\ifmmode~\breve{g}\else \u{g}\fi{}lu, Cederwall, Hadinia, Andgren,
  B\"ack, Grahn, Greenlees, Jakobsson, Johnson et~al.}}]{sandzelius09}
\bibinfo{author}{\bibfnamefont{M.}~\bibnamefont{Sandzelius}},
  \bibinfo{author}{\bibfnamefont{E.}~\bibnamefont{Ganio\ifmmode~\breve{g}\else
  \u{g}\fi{}lu}}, \bibinfo{author}{\bibfnamefont{B.}~\bibnamefont{Cederwall}},
  \bibinfo{author}{\bibfnamefont{B.}~\bibnamefont{Hadinia}},
  \bibinfo{author}{\bibfnamefont{K.}~\bibnamefont{Andgren}},
  \bibinfo{author}{\bibfnamefont{T.}~\bibnamefont{B\"ack}},
  \bibinfo{author}{\bibfnamefont{T.}~\bibnamefont{Grahn}},
  \bibinfo{author}{\bibfnamefont{P.}~\bibnamefont{Greenlees}},
  \bibinfo{author}{\bibfnamefont{U.}~\bibnamefont{Jakobsson}},
  \bibinfo{author}{\bibfnamefont{A.}~\bibnamefont{Johnson}},
  \bibnamefont{et~al.}, \bibinfo{journal}{Phys. Rev. C}
  \textbf{\bibinfo{volume}{79}}, \bibinfo{pages}{064315}
  (\bibinfo{year}{2009}).

\bibitem[{\citenamefont{Page et~al.}(2011)\citenamefont{Page, Andreyev,
  Wiseman, Butler, Grahn, Greenlees, Herzberg, Huyse, Jones, Jones
  et~al.}}]{page11}
\bibinfo{author}{\bibfnamefont{R.~D.} \bibnamefont{Page}},
  \bibinfo{author}{\bibfnamefont{A.~N.} \bibnamefont{Andreyev}},
  \bibinfo{author}{\bibfnamefont{D.~R.} \bibnamefont{Wiseman}},
  \bibinfo{author}{\bibfnamefont{P.~A.} \bibnamefont{Butler}},
  \bibinfo{author}{\bibfnamefont{T.}~\bibnamefont{Grahn}},
  \bibinfo{author}{\bibfnamefont{P.~T.} \bibnamefont{Greenlees}},
  \bibinfo{author}{\bibfnamefont{R.-D.} \bibnamefont{Herzberg}},
  \bibinfo{author}{\bibfnamefont{M.}~\bibnamefont{Huyse}},
  \bibinfo{author}{\bibfnamefont{G.~D.} \bibnamefont{Jones}},
  \bibinfo{author}{\bibfnamefont{P.~M.} \bibnamefont{Jones}},
  \bibnamefont{et~al.}, \bibinfo{journal}{Phys. Rev. C}
  \textbf{\bibinfo{volume}{84}}, \bibinfo{pages}{034308}
  (\bibinfo{year}{2011}).

\bibitem[{\citenamefont{Ulm et~al.}(1986)\citenamefont{Ulm, Bhattacherjee,
  Dabkiewicz, Huber, Kluge, K^^c3^^83^^c2^^bchl, Lochmann, Otten, Wendt, Ahmad
  et~al.}}]{ulm86}
\bibinfo{author}{\bibfnamefont{G.}~\bibnamefont{Ulm}},
  \bibinfo{author}{\bibfnamefont{S.}~\bibnamefont{Bhattacherjee}},
  \bibinfo{author}{\bibfnamefont{P.}~\bibnamefont{Dabkiewicz}},
  \bibinfo{author}{\bibfnamefont{G.}~\bibnamefont{Huber}},
  \bibinfo{author}{\bibfnamefont{H.-J.} \bibnamefont{Kluge}},
  \bibinfo{author}{\bibfnamefont{T.}~\bibnamefont{K^^c3^^83^^c2^^bchl}},
  \bibinfo{author}{\bibfnamefont{H.}~\bibnamefont{Lochmann}},
  \bibinfo{author}{\bibfnamefont{E.-W.} \bibnamefont{Otten}},
  \bibinfo{author}{\bibfnamefont{K.}~\bibnamefont{Wendt}},
  \bibinfo{author}{\bibfnamefont{S.}~\bibnamefont{Ahmad}},
  \bibnamefont{et~al.}, \bibinfo{journal}{Z. Phys. A}
  \textbf{\bibinfo{volume}{325}} (\bibinfo{year}{1986}).

\bibitem[{\citenamefont{Grahn et~al.}(2009)\citenamefont{Grahn, Petts, Scheck,
  Butler, Dewald, Hornillos, Greenlees, G\"orgen, Helariutta, Jolie
  et~al.}}]{grahn09}
\bibinfo{author}{\bibfnamefont{T.}~\bibnamefont{Grahn}},
  \bibinfo{author}{\bibfnamefont{A.}~\bibnamefont{Petts}},
  \bibinfo{author}{\bibfnamefont{M.}~\bibnamefont{Scheck}},
  \bibinfo{author}{\bibfnamefont{P.~A.} \bibnamefont{Butler}},
  \bibinfo{author}{\bibfnamefont{A.}~\bibnamefont{Dewald}},
  \bibinfo{author}{\bibfnamefont{M.~B.~G.} \bibnamefont{Hornillos}},
  \bibinfo{author}{\bibfnamefont{P.~T.} \bibnamefont{Greenlees}},
  \bibinfo{author}{\bibfnamefont{A.}~\bibnamefont{G\"orgen}},
  \bibinfo{author}{\bibfnamefont{K.}~\bibnamefont{Helariutta}},
  \bibinfo{author}{\bibfnamefont{J.}~\bibnamefont{Jolie}},
  \bibnamefont{et~al.}, \bibinfo{journal}{Phys. Rev. C}
  \textbf{\bibinfo{volume}{80}}, \bibinfo{pages}{014324}
  (\bibinfo{year}{2009}).

\bibitem[{\citenamefont{Warner and Casten}(1983)}]{War83}
\bibinfo{author}{\bibfnamefont{D.~D.} \bibnamefont{Warner}} \bibnamefont{and}
  \bibinfo{author}{\bibfnamefont{R.~F.} \bibnamefont{Casten}},
  \bibinfo{journal}{Phys. Rev. C} \textbf{\bibinfo{volume}{28}},
  \bibinfo{pages}{1798} (\bibinfo{year}{1983}).

\bibitem[{\citenamefont{Ma et~al.}(1993)\citenamefont{Ma, Hamilton, Ramayya,
  Chaturvedi, Deng, Gao, Jiang, Kormicki, Zhao, Johnson et~al.}}]{ma93}
\bibinfo{author}{\bibfnamefont{W.~C.} \bibnamefont{Ma}},
  \bibinfo{author}{\bibfnamefont{J.~H.} \bibnamefont{Hamilton}},
  \bibinfo{author}{\bibfnamefont{A.~V.} \bibnamefont{Ramayya}},
  \bibinfo{author}{\bibfnamefont{L.}~\bibnamefont{Chaturvedi}},
  \bibinfo{author}{\bibfnamefont{J.~K.} \bibnamefont{Deng}},
  \bibinfo{author}{\bibfnamefont{W.~B.} \bibnamefont{Gao}},
  \bibinfo{author}{\bibfnamefont{Y.~R.} \bibnamefont{Jiang}},
  \bibinfo{author}{\bibfnamefont{J.}~\bibnamefont{Kormicki}},
  \bibinfo{author}{\bibfnamefont{X.~W.} \bibnamefont{Zhao}},
  \bibinfo{author}{\bibfnamefont{N.~R.} \bibnamefont{Johnson}},
  \bibnamefont{et~al.}, \bibinfo{journal}{Phys. Rev. C}
  \textbf{\bibinfo{volume}{47}}, \bibinfo{pages}{R5} (\bibinfo{year}{1993}).

\bibitem[{\citenamefont{Davydov and Filippov}(1958)}]{Davydov58}
\bibinfo{author}{\bibfnamefont{A.~S.} \bibnamefont{Davydov}} \bibnamefont{and}
  \bibinfo{author}{\bibfnamefont{G.~F.} \bibnamefont{Filippov}},
  \bibinfo{journal}{Nucl. Phys.} \textbf{\bibinfo{volume}{8}},
  \bibinfo{pages}{237 } (\bibinfo{year}{1958}).

\bibitem[{\citenamefont{Chabanat et~al.}(1998)\citenamefont{Chabanat, Bonche,
  Haensel, Meyer, and Schaeffer}}]{SLy}
\bibinfo{author}{\bibfnamefont{E.}~\bibnamefont{Chabanat}},
  \bibinfo{author}{\bibfnamefont{P.}~\bibnamefont{Bonche}},
  \bibinfo{author}{\bibfnamefont{P.}~\bibnamefont{Haensel}},
  \bibinfo{author}{\bibfnamefont{J.}~\bibnamefont{Meyer}}, \bibnamefont{and}
  \bibinfo{author}{\bibfnamefont{R.}~\bibnamefont{Schaeffer}},
  \bibinfo{journal}{Nucl. Phys. A} \textbf{\bibinfo{volume}{635}},
  \bibinfo{pages}{231 } (\bibinfo{year}{1998}).

\bibitem[{\citenamefont{JOSHI et~al.}(1994)\citenamefont{JOSHI, ZGANJAR,
  RUPNIK, ROBINSON, MANTICA, CARTER, KORMICKI, GILL, WALTERS, BINGHAM
  et~al.}}]{joshi94}
\bibinfo{author}{\bibfnamefont{P.}~\bibnamefont{JOSHI}},
  \bibinfo{author}{\bibfnamefont{E.}~\bibnamefont{ZGANJAR}},
  \bibinfo{author}{\bibfnamefont{D.}~\bibnamefont{RUPNIK}},
  \bibinfo{author}{\bibfnamefont{S.}~\bibnamefont{ROBINSON}},
  \bibinfo{author}{\bibfnamefont{P.}~\bibnamefont{MANTICA}},
  \bibinfo{author}{\bibfnamefont{H.}~\bibnamefont{CARTER}},
  \bibinfo{author}{\bibfnamefont{J.}~\bibnamefont{KORMICKI}},
  \bibinfo{author}{\bibfnamefont{R.}~\bibnamefont{GILL}},
  \bibinfo{author}{\bibfnamefont{W.}~\bibnamefont{WALTERS}},
  \bibinfo{author}{\bibfnamefont{C.}~\bibnamefont{BINGHAM}},
  \bibnamefont{et~al.}, \bibinfo{journal}{International Journal of Modern
  Physics E} \textbf{\bibinfo{volume}{03}}, \bibinfo{pages}{757}
  (\bibinfo{year}{1994}).

\bibitem[{\citenamefont{Scheck et~al.}(2010)\citenamefont{Scheck, Grahn, Petts,
  Butler, Dewald, Gaffney, Hornillos, G\"orgen, Greenlees, Helariutta
  et~al.}}]{scheck10}
\bibinfo{author}{\bibfnamefont{M.}~\bibnamefont{Scheck}},
  \bibinfo{author}{\bibfnamefont{T.}~\bibnamefont{Grahn}},
  \bibinfo{author}{\bibfnamefont{A.}~\bibnamefont{Petts}},
  \bibinfo{author}{\bibfnamefont{P.~A.} \bibnamefont{Butler}},
  \bibinfo{author}{\bibfnamefont{A.}~\bibnamefont{Dewald}},
  \bibinfo{author}{\bibfnamefont{L.~P.} \bibnamefont{Gaffney}},
  \bibinfo{author}{\bibfnamefont{M.~B.~G.} \bibnamefont{Hornillos}},
  \bibinfo{author}{\bibfnamefont{A.}~\bibnamefont{G\"orgen}},
  \bibinfo{author}{\bibfnamefont{P.~T.} \bibnamefont{Greenlees}},
  \bibinfo{author}{\bibfnamefont{K.}~\bibnamefont{Helariutta}},
  \bibnamefont{et~al.}, \bibinfo{journal}{Phys. Rev. C}
  \textbf{\bibinfo{volume}{81}}, \bibinfo{pages}{014310}
  (\bibinfo{year}{2010}).

\bibitem[{\citenamefont{Proetel et~al.}(1974)\citenamefont{Proetel, Diamond,
  and Stephens}}]{proetel74}
\bibinfo{author}{\bibfnamefont{D.}~\bibnamefont{Proetel}},
  \bibinfo{author}{\bibfnamefont{R.}~\bibnamefont{Diamond}}, \bibnamefont{and}
  \bibinfo{author}{\bibfnamefont{F.}~\bibnamefont{Stephens}},
  \bibinfo{journal}{Phys. Lett. B} \textbf{\bibinfo{volume}{48}},
  \bibinfo{pages}{102 } (\bibinfo{year}{1974}).

\bibitem[{\citenamefont{Ma et~al.}(1986)\citenamefont{Ma, Ramayya, Hamilton,
  Robinson, Cole, Zganjar, Spejewski, Bengtsson, Nazarewicz, and Zhang}}]{ma86}
\bibinfo{author}{\bibfnamefont{W.}~\bibnamefont{Ma}},
  \bibinfo{author}{\bibfnamefont{A.}~\bibnamefont{Ramayya}},
  \bibinfo{author}{\bibfnamefont{J.}~\bibnamefont{Hamilton}},
  \bibinfo{author}{\bibfnamefont{S.}~\bibnamefont{Robinson}},
  \bibinfo{author}{\bibfnamefont{J.}~\bibnamefont{Cole}},
  \bibinfo{author}{\bibfnamefont{E.}~\bibnamefont{Zganjar}},
  \bibinfo{author}{\bibfnamefont{E.}~\bibnamefont{Spejewski}},
  \bibinfo{author}{\bibfnamefont{R.}~\bibnamefont{Bengtsson}},
  \bibinfo{author}{\bibfnamefont{W.}~\bibnamefont{Nazarewicz}},
  \bibnamefont{and} \bibinfo{author}{\bibfnamefont{J.-Y.} \bibnamefont{Zhang}},
  \bibinfo{journal}{Phys. Lett. B} \textbf{\bibinfo{volume}{167}},
  \bibinfo{pages}{277 } (\bibinfo{year}{1986}).

\bibitem[{\citenamefont{Angeli}(2004)}]{angeli04}
\bibinfo{author}{\bibfnamefont{I.}~\bibnamefont{Angeli}}, \bibinfo{journal}{At.
  Data and Nucl. Data Tables} \textbf{\bibinfo{volume}{87}},
  \bibinfo{pages}{185 } (\bibinfo{year}{2004}).

\bibitem[{\citenamefont{{WWW Table of Atomic Masses}}()}]{mass_data}
\bibinfo{author}{\bibnamefont{{WWW Table of Atomic Masses}}},
  \bibinfo{howpublished}{\url{http://ie.lbl.gov/toi2003/MassSearch.asp}}.

\bibitem[{\citenamefont{Kib\'edi and Spear}(2005)}]{kibedi05}
\bibinfo{author}{\bibfnamefont{T.}~\bibnamefont{Kib\'edi}} \bibnamefont{and}
  \bibinfo{author}{\bibfnamefont{R.}~\bibnamefont{Spear}},
  \bibinfo{journal}{At. Data and Nucl. Data Tables}
  \textbf{\bibinfo{volume}{89}}, \bibinfo{pages}{77 } (\bibinfo{year}{2005}).

\end{thebibliography}

\end{document}